\pdfoutput=1
\documentclass[traditabstract]{aa} 
\usepackage{graphicx}
\usepackage{natbib}
\usepackage{epsfig}
\usepackage{caption}
\usepackage{subcaption}
\usepackage{wasysym}
\usepackage{afterpage}
\usepackage{float}
\usepackage{amsfonts}
\usepackage{enumerate}
\usepackage{txfonts}
\def\ms{\,m\,s$^{-1}$}         
\def\kms{\hbox{\,km\,s$^{-1}$}}       
\def\vsini{\hbox{$v$\,sin\,$i_\star$}}      
\newcommand{\Msun}{$M_{\odot}$}             
\def\Rsun{\hbox{$R_{\odot}$}}

\def \e{\times 10^}
\newcommand{\vrad}{$v_{\rm rad}$} 
\newcommand{\teff}{$T_{\rm eff}$}
\newcommand{\logg}{$\log g$}

\newcommand{\vmac}{$v_{\rm macro}$}
\newcommand{\feh}{[Fe/H]}
\newcommand{\MJ}{{\it M}$_{\rm J}$}

\newcommand{\RJ}{{\it R}$_{\rm J}$}

\newcommand{\sn}{S/N}

\begin{document}

\title{SOPHIE velocimetry of Kepler transit candidates  \\
       XIV. A joint photometric, spectroscopic, and dynamical analysis of the Kepler-117 system\thanks{Appendix A is available in electronic form at \texttt{http://www.aanda.org}.}$^,$\thanks{Radial velocity tables are only available at the CDS via anonymous ftp to \texttt{cdsarc.u-strasbg.fr (130.79.128.5)} or via \texttt{http://cdsarc.u-strasbg.fr/viz-bin/qcat?J/A+A/573/A124}.}}

\author{
G. Bruno\inst{\ref{lam}}
\and J.-M. Almenara\inst{\ref{lam}} 
\and S. C. C. Barros\inst{\ref{lam}}
\and A. Santerne\inst{\ref{lam},\ref{caup}, \ref{porto2}}
\and R. F. Diaz\inst{\ref{geneve}}
\and M. Deleuil\inst{\ref{lam}}
\and C. Damiani\inst{\ref{lam}}
\and A. S. Bonomo \inst{\ref{inaf}}
\and I. Boisse\inst{\ref{lam}}
\and F. Bouchy\inst{\ref{lam}}
\and G. H\'ebrard\inst{\ref{ohp},\ref{iap}}
\and G. Montagnier\inst{\ref{ohp},\ref{iap}}
        }

\institute{
Aix Marseille Universit\'e, CNRS, LAM (Laboratoire d'Astrophysique de Marseille) UMR 7326, 13388, Marseille, France\label{lam}
\and Centro de Astrof\'isica, Universidade do Porto, Rua das Estrelas, 4150-762 Porto, Portugal\label{caup}
\and Instituto de Astrof\'isica e Ci\^{e}ncias do Espa\c co, Universidade do Porto, CAUP, Rua das Estrelas, PT4150-762 Porto, Portugal\label{porto2}
\and Observatoire Astronomique de l'Universit\'e de Gen\`eve, 51 chemin des Maillettes, 1290 Versoix, Switzerland\label{geneve}
\and INAF - Osservatorio Astrofisico di Torino, via Osservatorio 20, 10025 Pino Torinese, Italy \label{inaf}
\and Observatoire de Haute Provence, 04670 Saint Michel l'Observatoire, France\label{ohp}
\and Institut d'Astrophysique de Paris, UMR7095 CNRS, Universit\'e Pierre \& Marie Curie, 98bis boulevard Arago, 75014 Paris, France\label{iap} 
          }
\date{Received 11 July 2014 / Accepted 8 November 2014}

 
  \abstract
   {As part of our follow-up campaign of Kepler planets, we observed Kepler-117 with the SOPHIE spectrograph at the Observatoire de Haute-Provence. This F8-type star hosts two transiting planets in non-resonant orbits. The planets, Kepler-117 b and c, have orbital periods $\simeq 18.8$ and $\simeq 50.8$ days, and show transit-timing variations (TTVs) of several minutes. We performed a combined Markov chain Monte Carlo (MCMC) fit on transits, radial velocities, and stellar parameters to constrain the characteristics of the system. We included the fit of the TTVs in the MCMC by modeling them with dynamical simulations. In this way, consistent posterior distributions were drawn for the system parameters. According to our analysis, planets b and c have notably different masses ($0.094 \pm 0.033$ and $1.84 \pm 0.18$ \MJ) and low orbital eccentricities ($0.0493 \pm 0.0062$ and $0.0323 \pm 0.0033$). The uncertainties on the derived parameters are strongly reduced if the fit of the TTVs is included in the combined MCMC. The TTVs allow measuring the mass of planet b, although its radial velocity amplitude is poorly constrained. Finally, we checked that the best solution is dynamically stable.}


\titlerunning{Kepler-117}
\authorrunning{G. Bruno et al.}

   \maketitle


\section{Introduction}
In the past few years, the number of known multiple planet systems detected by the Kepler space telescope has enormously increased. Multiple transiting planet systems have a low false-positive probability: using conservative hypotheses, \cite{lissauer2012} estimated a 1.12\% probability of observing two false positives in the same system and a 2.25\% probability for a system to host a planet and show the features of a false positive at the same time. Their estimation was based on the assumptions that false positives are randomly distributed among the Kepler targets and that there is no correlation between the probability of a target to host one or more detectable planets and display false positives.\\
At the time of writing, Kepler has detected 469 multiple planet systems.\footnote{http://exoplanet.eu/.} 
Among them, Kepler-117 (also named KOI-209) hosts the two transiting planets Kepler-117 b and Kepler-117 c. These planets were presented by \cite{borucki2011} as candidates and validated by \cite{rowe2014} with a confidence level of more than 99\%, while radial velocity observations were still unavailable. After subjecting Kepler-117 b and c to various false-positive identification criteria, \citeauthor{rowe2014} used the statistical framework of \cite{lissauer2012} (further refined in \citealt{lissauer2014}) to promote them to \textit{bona fide} exoplanets. Kepler-117 b and c were found to have orbital periods $\simeq 18.8$ and $\simeq 50.8$ days and radii $\simeq 0.72$ and $\simeq 1.04$ \RJ.\\
Multiple planet systems offer insights on their dynamical history \citep[e.g.,][]{batygin2013} and can show transit-timing variations \citep[TTVs;][]{agol2005}, especially if the planets are in mean motion resonance. TTVs are a powerful tool for detecting non-transiting planets, and for the determination of planetary masses \citep{holman2010,nesvorny2013,barros2014,dawson2014} and can be a tracer of stellar activity, as well \citep{barros2013,oshagh2013}. Moreover, non-detected TTVs can cause an underestimation of the uncertainty on the stellar density derived from the photometry \citep{kipping2014}.\\
Using only the first quarter of the Kepler photometric data, \cite{steffen2010} predicted TTVs to be observable for the Kepler-117 system. At that time, the photometric time coverage was not sufficient to allow a verification. The TTVs were later confirmed by \cite{mazeh2013}. According to these authors, the ratio between the periodic modulation of the TTVs of the inner planet (b) and the orbital period of the outer one (c), $P_{\rm TTVs, b}/P_c$, is $\simeq 0.997$. This ratio is the closest to 1 among the Kepler candidate two-planets systems with TTVs presented in that paper. The similarity between the two periodicities is a strong indication that the two bodies are in the same system and thus is another argument for the validation of Kepler-117 b and c.\\ 
In this paper, we  included the information from the TTVs in the combined fit of the system parameters together with the photometry and radial velocities we acquired during our observation campaigns with the SOPHIE spectrograph at the Observatoire de Haute-Provence. By fully exploiting the data, we obtained a precise measure of the masses and radii of the planets. In Sect. \ref{data}, the data acquisition and reduction is discussed. In Sect. \ref{host}, we describe the treatment of the stellar spectra, and in Sect. \ref{analysis} we report on the stellar activity, the measurement of the TTVs, and the joint Bayesian fit of the system parameters. In Sect. \ref{disc}, the results are discussed. The implications and conclusions are given in Sect. \ref{conclusions}.

\section{Observations and data reduction}\label{data}
\subsection{Kepler photometric observations}
\label{photom}

Kepler-117 was observed by the Kepler space telescope from quarter 1 to 17 between May 2009 and May 2013. The first three quarters were covered by a sampling of 29.4 minutes (long-cadence data, LC), the following were sampled every 58.5 seconds (short-cadence data, SC). We chose to use the LC data only for quarters from 1 to 3, and relied on the SC data for the others. The light curves, already reduced by the Kepler pipeline \citep{jenkins2010}, are publicly available on the Mikulski Archive for Space Telescopes (MAST).\footnote{http://archive.stsci.edu/index.html.} We made use of the light curves corrected by the Presearch Data Conditioning (PDC) module, available in the light curve \texttt{fits} file. \\ 
For all the quarters and for both the LC and the SC data, the dispersion of the contamination of nearby stars, corrected for by the pipeline, is lower than 1\%. The contamination value, then, was fixed in the following combined analysis (Sect. \ref{modelling}).\\
We isolated the photometric signal around every transit using a preliminary estimate of the ephemeris, following \cite{rowe2014}. The transits of the two planets sometimes superpose because of their different periods ($\simeq 18.8$ days and $\simeq 50.8$ days). We discarded these overlapping transits because the software we used to fit the data sets (Sect. \ref{modelling}) does not yet include this modeling. No secondary eclipse was found, as expected from the relatively long periods.\\
We normalized the transits by fitting a second-order polynomial to the flux outside of the transits and rejected the outliers through a $3\sigma$ clipping.

\subsection{Spectroscopic observations}
\label{spectro}

\begin{figure}[htbp]
\includegraphics[scale = 0.45]{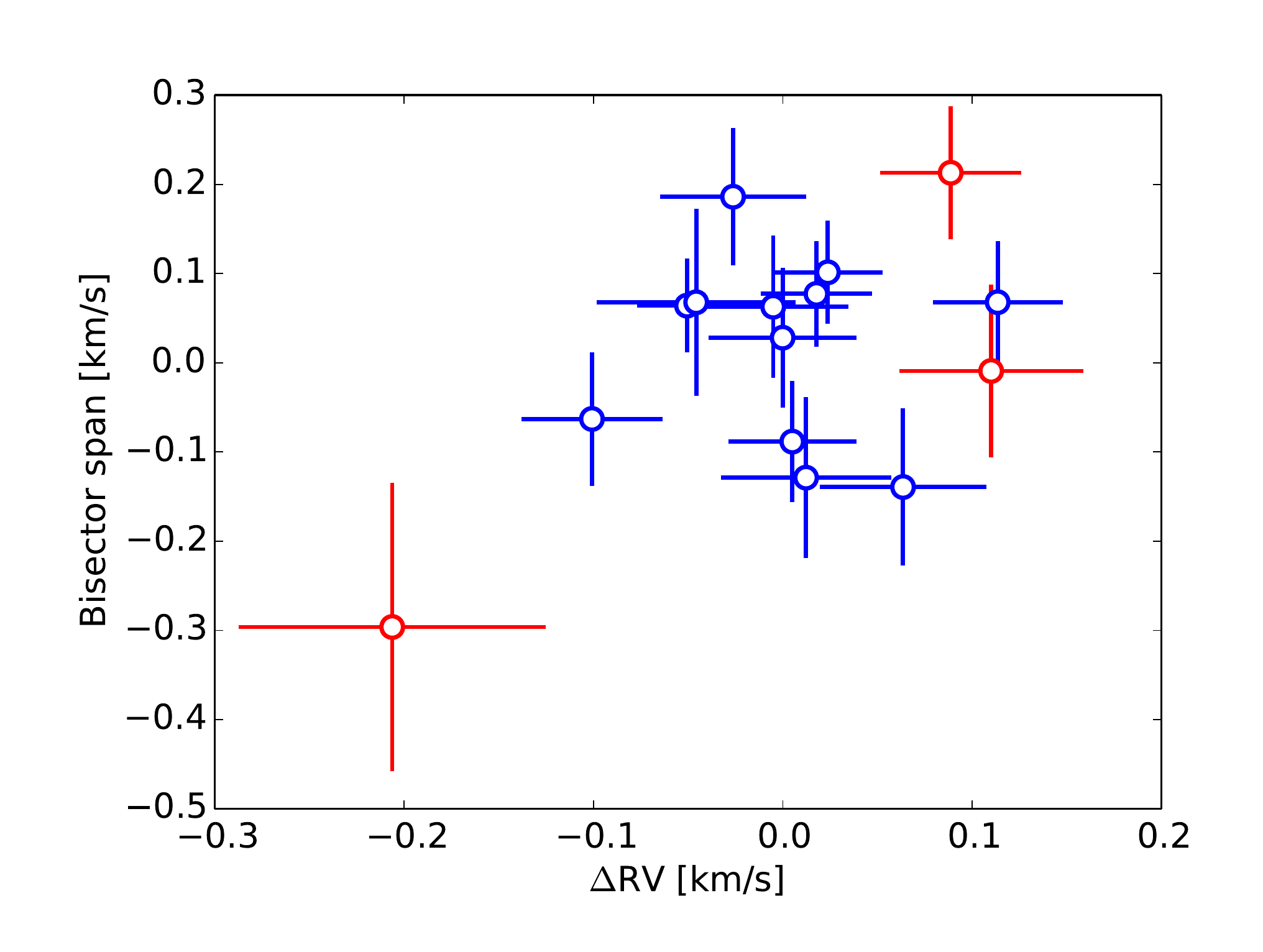}
\caption{Bisector span of the CCF plotted with respect to the radial velocity measurements (the mean RV has been subtracted). The red points indicate contamination by the Moon. The uncertainty on the bisector span of each point is twice the uncertainty on the RV for that point.}
\label{biss}
\end{figure}

Kepler-117 is part of our follow-up program of Kepler candidates \citep{bouchy2011,santerne2012}. We acquired 15 spectra of this star during two observing seasons, between July 2012 and November 2013, using the SOPHIE spectrograph at the 1.93 m telescope of the Observatoire de Haute-Provence \citep{perruchot2008,bouchy2013}. The instrument was set in high-efficiency mode, with a spectral resolution $\lambda/\Delta \lambda \sim 38000$. The exposures lasted from 1200 to 3600 s for an \sn\ per pixel at 550 nm between 9 and 17. The spectra were reduced using the SOPHIE pipeline \citep{bouchy2009}. The radial velocities (RVs) and their uncertainties were obtained through a Gaussian fit of the cross-correlation function (CCF) with numerical masks corresponding to the F0, G2, and K5 spectral types. The final RVs were measured with the G2 mask because the spectral analysis showed Kepler-117 to be close to a G star (i.e., \citealp{rowe2014}, verified in Sect. \ref{host}). However, using different masks to compute the CCF did not result in a systematic difference between the RVs. The RV reference star HD185144 \citep{howard2010,bouchy2013,santerne2014} was used to correct the RVs by between $\sim 5$ and $\sim 30$ \ms. Three spectra were affected by the moonlight: we corrected them for the RV of the Moon, as discussed in \cite{baranne1996}, \cite{pollacco2008}, and \cite{hebrard2008}. The charge transfer inefficiency effect was corrected for using the prescription of \cite{santerne2012}. The first three echelle orders at the blue edge of the spectrum were not used to calculate of the RVs because their low \sn\ degrades the precision of the measurements.\\
We rejected the point at $\rm{BJD} = 2456551.49295$ because of its low \sn\ (9 at 550 nm, the lowest of all the set).\\
We checked for linear correlations between the bisector span of the CCF and the RVs, following \cite{queloz2001} (Fig. \ref{biss}). If linear correlations are observed, the planetary scenario is very likely to be rejected in favor of a blend. The Spearman-rank-order correlation coefficient between the bisector span and the RVs, excluding the points contaminated by the moonlight, is $-0.08 \pm 0.32$. The p-value for this coefficient, with the null hypothesis of no correlation, is 0.98. Similarly, the Spearman correlation coefficient between the full width at half maximum of the CCFs of the spectra and their respective RV is $0.27 \pm 0.30$, with a p-value of 0.39. The two diagnostics on the CCF are clearly compatible with the planetary scenario.

\begin{table}[htbp]
\begin{center}{
\caption{\label {TabRV}  Log of SOPHIE radial velocity observations. The points marked with \rightmoon\ in the date are contaminated by moonlight, while the one with $\dagger$ was discarded from the analysis (Sect. \ref{spectro}).}
\begin{tabular}{|c|c|c|c|c|}
\hline
\hline
Date  & BJD    & \vrad\  & $\sigma$\vrad  & \sn\ at \\
          &      - 2450000     & [\kms]    &  [\kms]           &  550 nm$^\ast$ \\ 
\hline
2012-07-15    &     6123.54450   &  -12.842 & 0.034 &   14.12     \\
2012-07-24    &     6133.44333   &  -12.932  & 0.029  &   15.38     \\
2012-08-13    &     6153.46918   &  -13.056 & 0.037 &   13.46     \\
2012-08-22    &     6161.50353   &   -12.892 & 0.044 &   13.87     \\
2012-09-09    &     6180.46953   &  -12.943 & 0.045  &   11.83     \\
2012-09-17    &     6188.38902   &  -12.961  & 0.040 &   14.57     \\
2012-10-13    &     6214.28684   &  -12.956 & 0.039  &   12.75     \\
2013-05-08    &     6420.57184   &  -12.951 & 0.034  &   13.27     \\
2013-08-01    &     6505.52985   &  -12.982 & 0.039  &   14.76     \\
2013-08-29 \rightmoon\    &     6533.52259   &  -12.846 &	0.049  &   11.67     \\
2013-09-16\rightmoon,$\dagger$    &     6551.49295   &  -13.162 & 0.081  &   9.27      \\
2013-09-23    &     6559.30819   &  -13.006 & 0.026  &   16.44     \\
2013-10-16 \rightmoon\  &     6582.31397   &  -12.867 &	0.037  &   16.83     \\
2013-10-27    &     6593.37762   &  -12.938 & 0.030  &   15.33     \\
2013-11-23    &     6620.24985   &  -13.001 & 0.053  &   12.47     \\
\hline
\end{tabular}}
\end{center}
\begin{list}{}{}
\item $^\ast$ Measured by the SOPHIE pipeline.
\end{list}
\end{table}

\section{Host star}
\label{host}
To analyze the stellar atmosphere we used only spectra that were not affected by moonlight. The only spectrum at  an $\rm{S/N}< 10$ at 550 nm was discarded as well. The remaining twelve spectra were shifted according to their measured RVs, corrected for the cosmic rays neglected by the SOPHIE pipeline, co-added, and normalized. The final \sn\ in the continuum, at 550 nm, is $\simeq 130$ per resolution element.\\ 
We measured the stellar effective temperature \teff, surface gravity \logg, metallicity \feh, and projected rotational velocity \vsini\ with the \texttt{VWA} software \citep[and references therein]{bruntt201023,bruntt2010c7}. This method is based on the fit of the metal spectral lines, especially the iron lines. The best parameters are those that minimize the correlation of the element abundances with the excitation potential and the equivalent width of the spectral lines. We obtained $\rm{T}_{\rm eff} = 6260 \pm 80$ K, $\log g = 4.40 \pm 0.11$, and $\rm{[Fe/H]} = 0.10 \pm 0.13$, appropriate of an F8V-type star.\\
The estimate of \logg\ was confirmed with the pressure-sensitive lines of CaI at 612.2 nm and the MgIb triplet. Finally, the couple \vsini\ and \vmac\ was jointly measured by fitting a rotational profile on a set of isolated spectral lines. This measure of \vsini\ ($6 \pm 2$ \kms) agrees with the one obtained with the fit of the CCF \citep{boisse2010}: $6.8 \pm 1.0$ \kms.\\

\begin{figure}[!htbp]
\centering
\includegraphics[scale = 0.65]{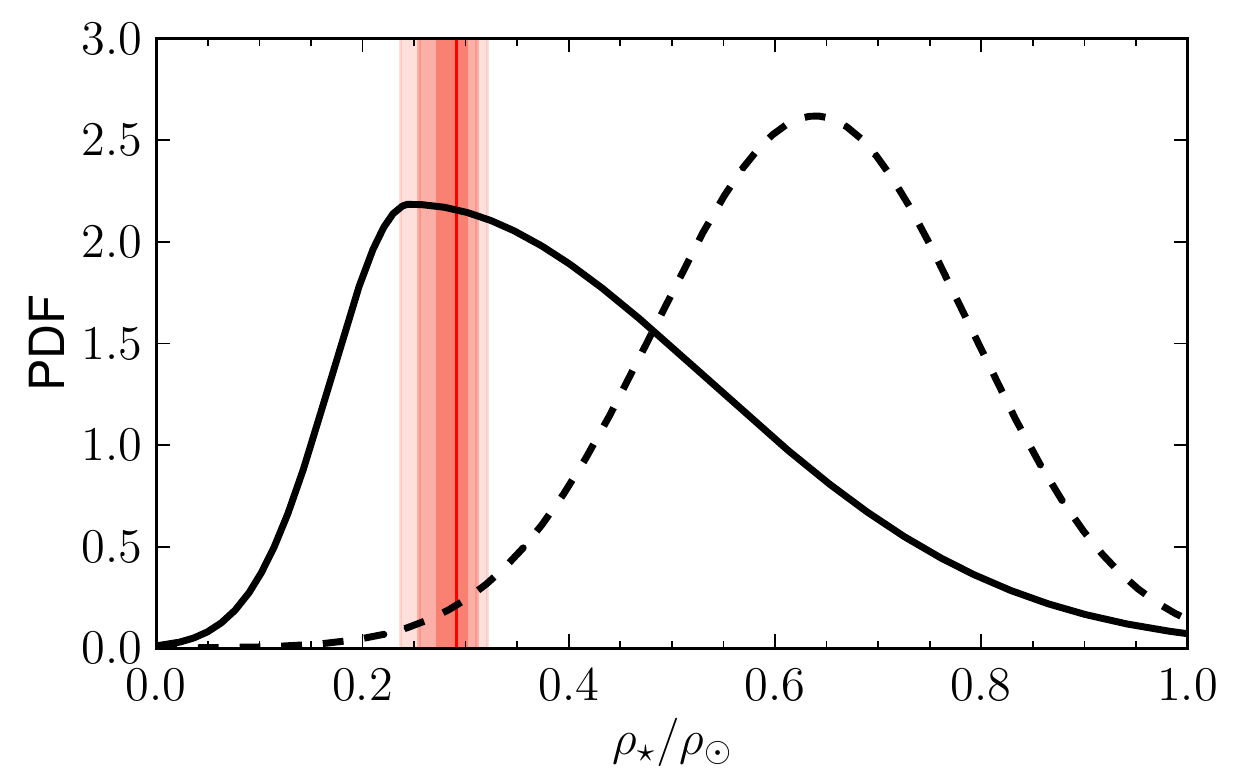}
\caption{Stellar densities derived from the spectroscopic parameters of the SOPHIE spectrum (dashed line) and the HIRES spectrum \citep{rowe2014} (continuous line). In red, the posterior distribution from the PASTIS analysis, shaded according to the 1-, 2-, and $3\sigma$ intervals.}
\label{compdens}
\end{figure}

A first combined fit of the data sets (Sect. \ref{modelling}) showed a $\sim 3\sigma$ difference between the spectroscopic \logg\ ($4.40 \pm 0.11$) and the one derived from the posteriors of the stellar parameters ($4.102\pm0.019$). We compared our spectroscopic parameters with the other published ones. \cite{everett2013} observed Kepler-117 at the National Optical Astronomy Observatory (NOAO) Mayall 4m telescope on Kitt Peak with the RCSpec long-slit spectrograph. They reported the result of two fits on the spectrum, obtaining $\log g = 4.26\pm 0.15$ and $4.65\pm 0.15$ (reaching, in this case, a parameter limit in their model). \cite{rowe2014}, instead, used the publicly available spectra recorded with the HIRES spectrograph at the Keck I telescope and found $\log g = 4.187 \pm 0.150$. This star has been observed with other telescopes as well, but the resulting parameters are published only in the Kepler Community Follow-up Observing Program (CFOP) online archive\footnote{https://cfop.ipac.caltech.edu/home/.} and not in the literature, so that we did not consider them. The \logg\ is known to be a problematic parameter to measure accurately and is correlated with \teff\ and \feh. In particular, a decrease in \logg\ is usually reflected by a decrease in \teff\ and in \feh. The complete set of the three parameters in the articles we referred to is reported in Table \ref{tabspec}. The stellar densities derived from the SOPHIE spectrum, the HIRES spectrum (both calculated with the Dartmouth tracks), and those from the TTVs are shown in Fig. \ref{compdens}.\\
We were unable to identify a problem in the SOPHIE spectra nor in our analysis method. We therefore chose to use the published combination of \teff, \logg\, and \feh\ whose \logg\ is the closest to our posterior, which converges to a sharp distribution even in the tail of the large spectroscopic prior of SOPHIE. The final values we adopted are those of \cite{rowe2014}, which were used as priors in the Bayesian analysis. 

\begin{table}[htbp]
\caption{\label {tabspec} Published spectroscopic parameters for Kepler-117 compared with those of this work.}
\begin{tabular}{|l|l|l|l|}
\hline\hline
Authors & \teff [K] & \logg & \feh \\
\hline
\cite{everett2013} (1) & $6185\pm 75$ & $4.25 \pm 0.15$ & $-0.04 \pm 0.10$ \\
\cite{everett2013} (2)$^\ast$ & $6316\pm 75$ & $4.65 \pm 0.15$ & $0.09 \pm 0.10$ \\
\cite{rowe2014}    & $6169 \pm 100$ & $4.187 \pm 0.150$ & $-0.04 \pm 0.10$ \\
This work          & $6260 \pm 80$ & $4.40 \pm 0.11$ & $0.10 \pm 0.13$\\
\hline
\end{tabular}
\begin{list}{}{}
\item $^{\ast}$ Two set of parameters are presented in \cite{everett2013}. The fit marked with (2) is reported to have reached a parameter limit in the models. 
\end{list}
\end{table}

\section{System analysis}\label{analysis}

\subsection{Stellar activity}\label{activity}
\begin{figure}[htbp]
\centering
\includegraphics[scale = 0.45]{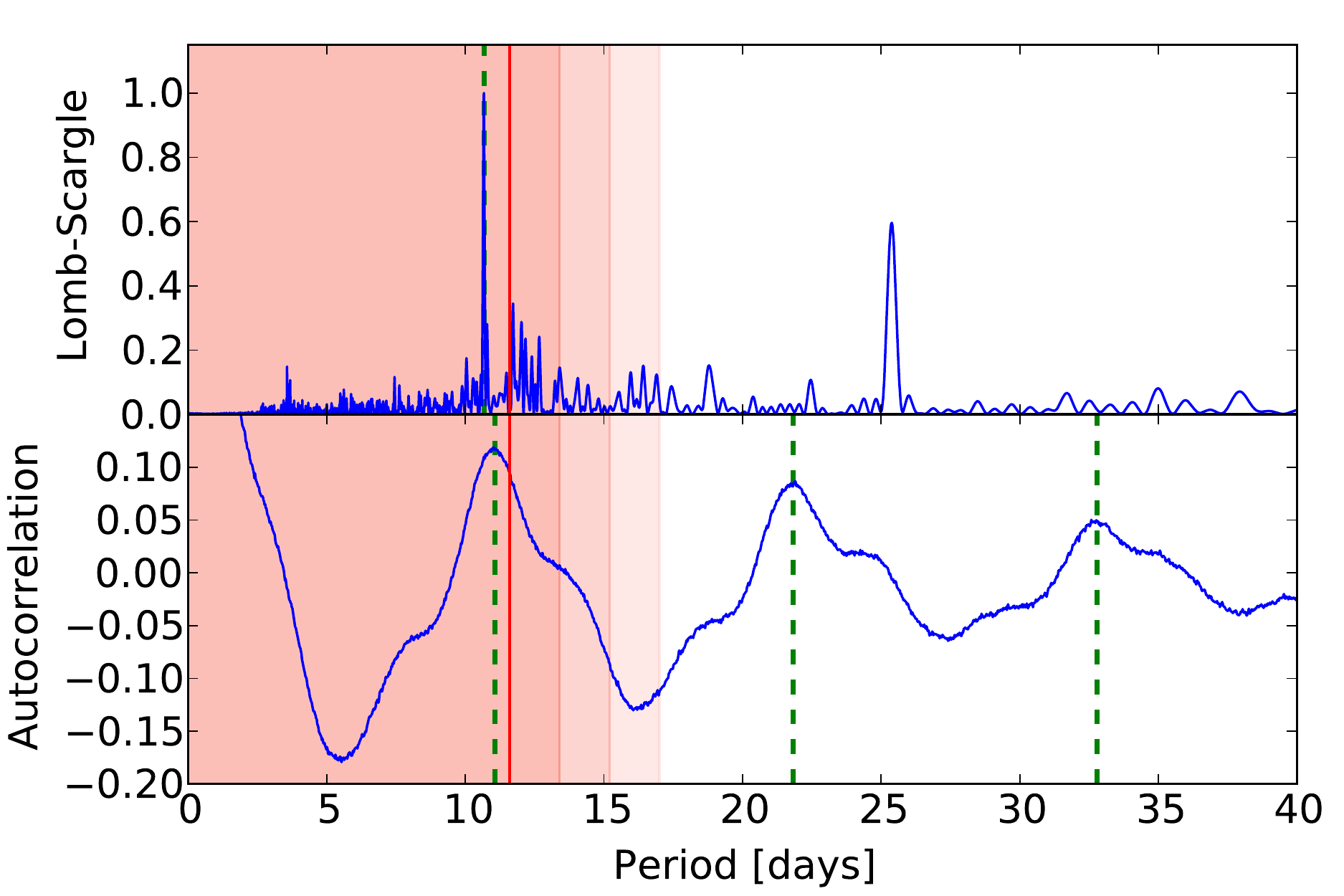}
\caption{\textit{Top panel}: Lomb-Scargle periodogram of the light curve after removing the transits. \textit{Bottom panel}: autocorrelation of the light curve. The green dotted lines indicate the Gaussian fit to the peaks. The red line corresponds to the maximum rotation period of the star deduced by the \vsini and the stellar radius. The red shadowed regions highlight the 1-, 2-, and $3\sigma$ confidence intervals for the rotation period.}
\label{perio}
\end{figure}

The light curve shows small periodic variations, arguably due to starspots. 
To identify the periodicities, we removed the transit features and computed the Lomb-Scargle periodogram (LSP: \citealp{press1989}) of the light curve (Fig. \ref{perio}), finding a peak at $10.668 \pm 0.028$ days. The uncertainty is underestimated because it does not take into account the position of the spots on the stellar surface and the differential rotation. The $\simeq 11$-day periodicity was also isolated by the autocorrelation of the light curve (Fig. \ref{perio}, bottom panel), precisely, by the main peak at $11.1 \pm 1.4$ days and the first two multiples at $21.8 \pm 1.7$ and $32.8 \pm 1.9$ days. The measurements agree with the stellar rotation period measured from the \vsini\ (Sect. \ref{host}) and the stellar radius $R_\star$ (Sect. \ref{modelling}), assuming that the rotation axis is perpendicular to the line of sight ($11.6\pm1.8$ days). In conclusion, the peaks in the periodogram and the autocorrelation function can be considered as representative of the rotation of the host 
star, for which we conservatively adopted the photometric value with the largest uncertainty, that is, $P_\star = 11.1 \pm 1.4$ days. Comparing this with \vsini\ shows that the stellar inclination is compatible with $90^\circ$.

\begin{figure*}[htbp]
\centering
\includegraphics[scale = 0.45]{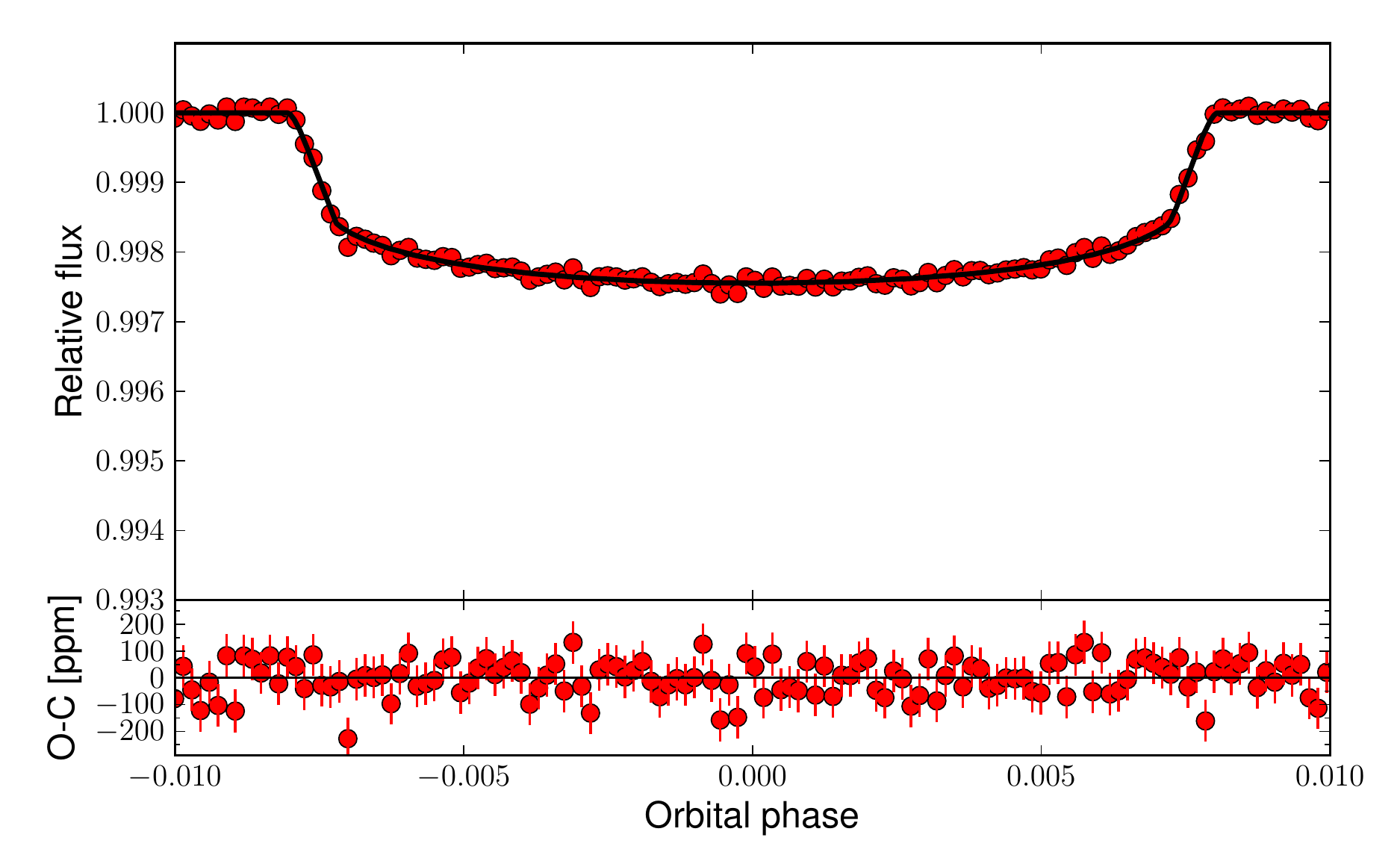}
\includegraphics[scale = 0.45]{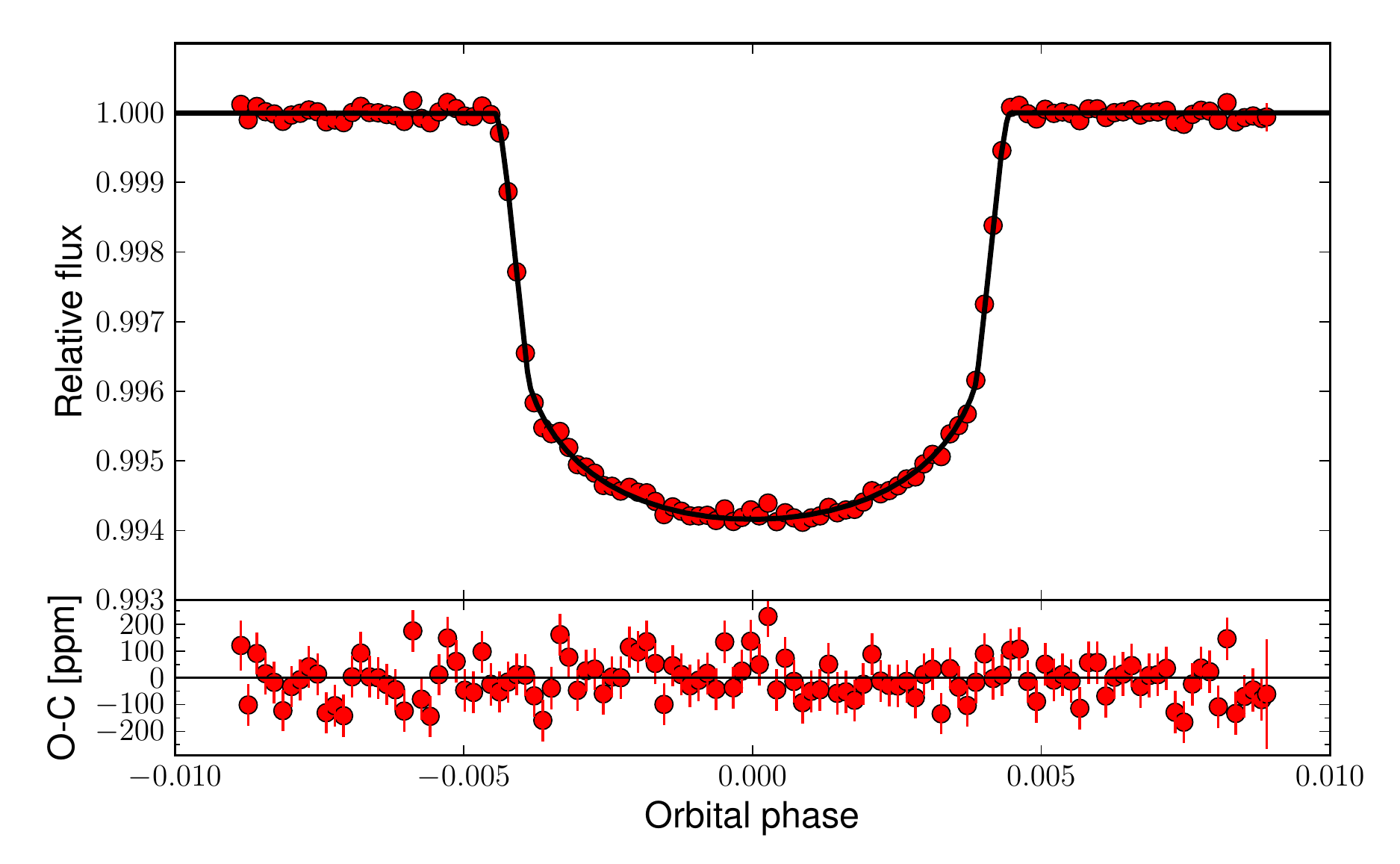}
\includegraphics[scale = 0.45]{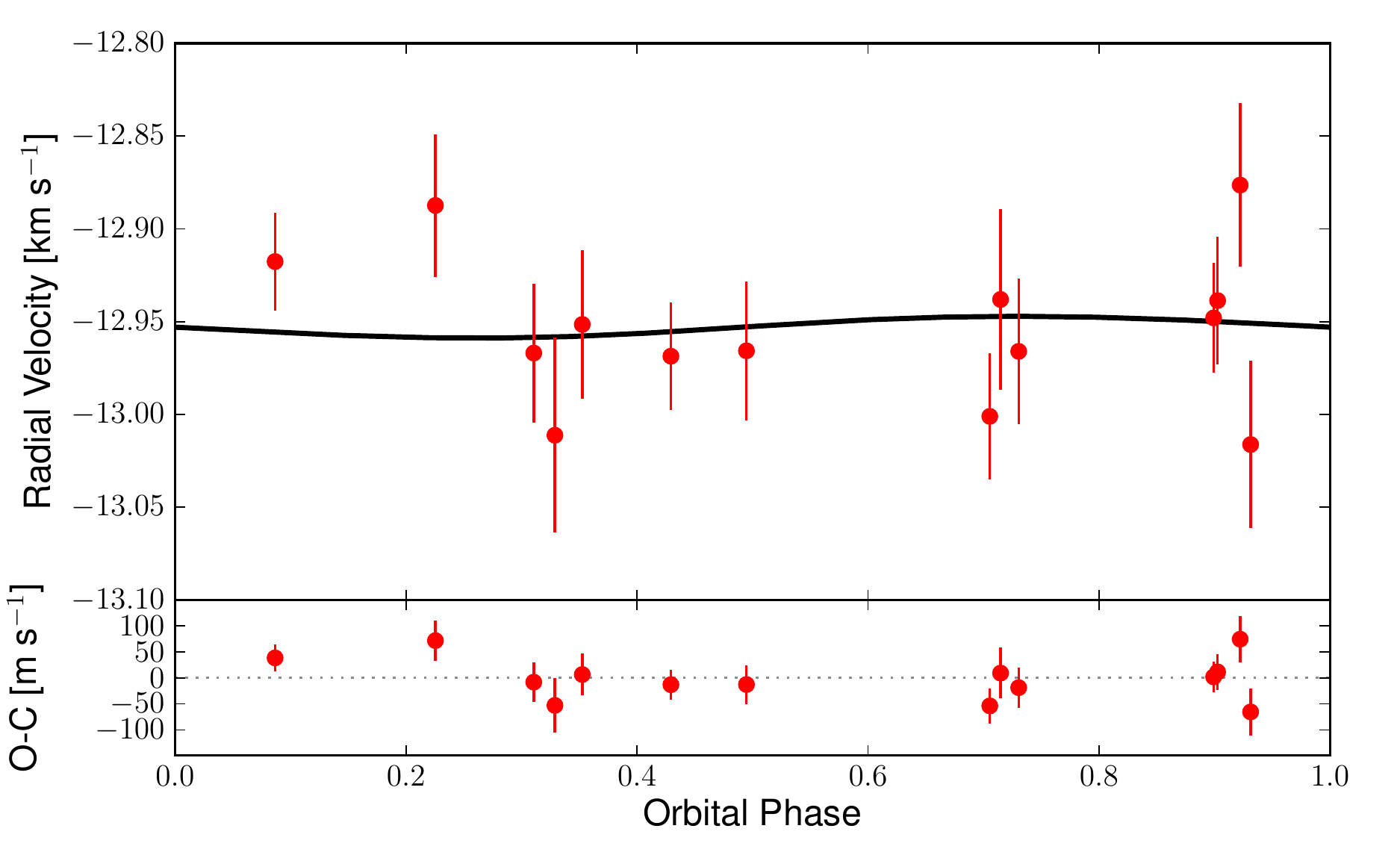}
\includegraphics[scale = 0.45]{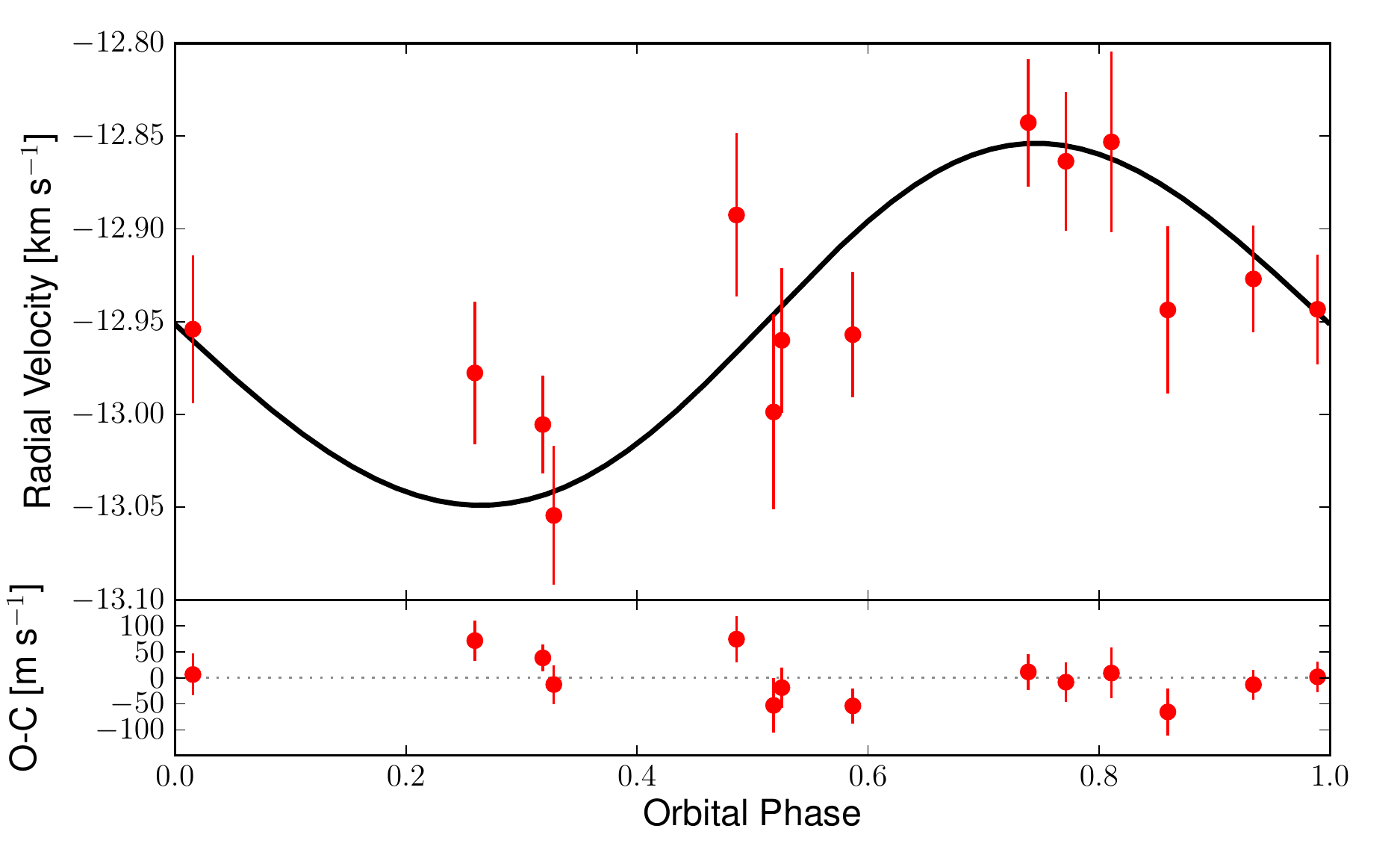}
\includegraphics[scale = 0.45]{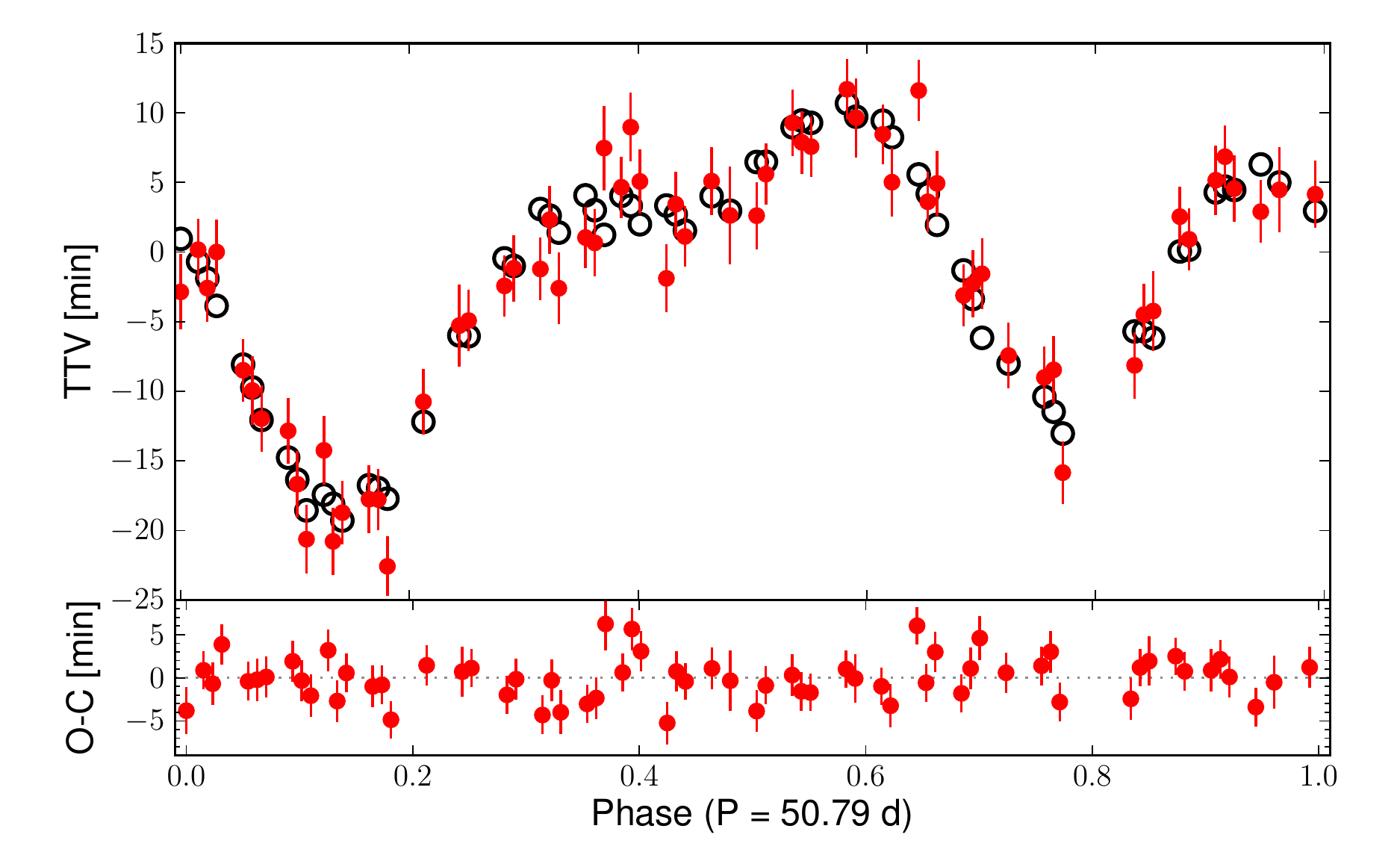}
\includegraphics[scale = 0.45]{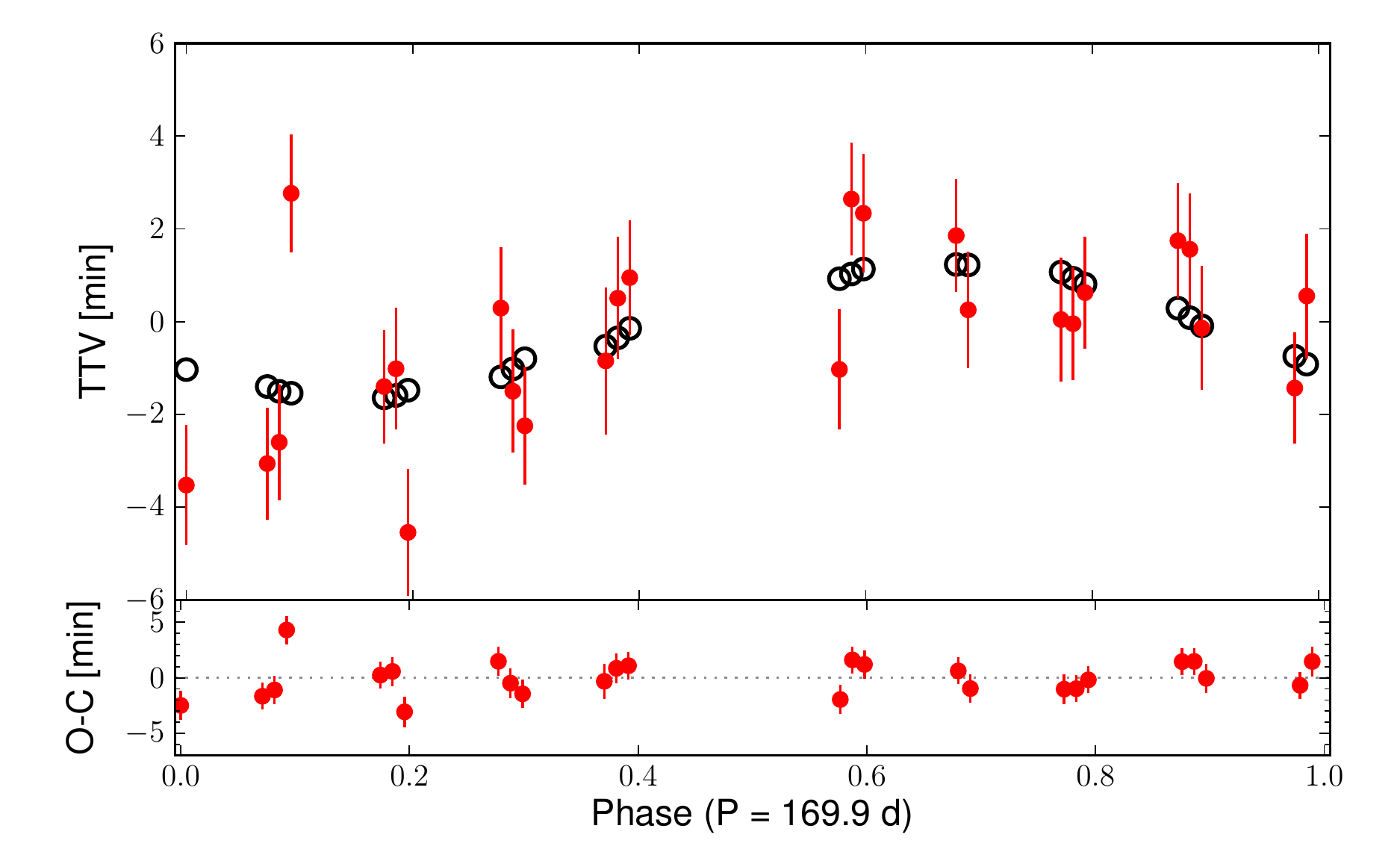}
\caption{\textit{Top}: phase-folded plot of the best transit model of planet b (left) and c (right), over the SC data. In black the model, in red the data binned every hundredth of orbital phase. \textit{Center}: the same for the radial velocities. \textit{Bottom}: The TTVs of planet b folded at the orbital period of planet c (left) and those of planet c folded at the first peak of its Lomb-Scargle periodogram (right, Sect. \ref{modu}). For each plot, the lower panel shows the residuals as observed minus calculated ($O-C$) points.}
\label{modeldata}
\end{figure*}

\subsection{Combined fit of the system parameters}
We performed a combined Bayesian analysis of photometry, radial velocities, stellar parameters, and TTVs. We begin by describing the measurement and modeling of the TTVs and then present the Bayesian analysis.

\subsubsection{Transit time variations}\label{measttvs}
After reducing the photometric data (Sect. \ref{photom}), we fitted the transit times with a procedure similar to the one discussed in \cite{barros2011}. All the transits were fitted simultaneously to constrain the shape parameters, that is, the normalized separation of the planet $a/R_\star$, the ratio of planet-to-star radius $R_p /R_\star$, and the orbital inclination $i$. For each transit, the primary transit epoch $T_0$ and three normalization parameters were also fitted to account for a quadratic trend with time. The derived transit times for each planet are given in Tables \ref{ttvb} and \ref{ttvc}. After removing a linear ephemeris, the transit times of the two planets showed significant TTVs. The TTVs exhibited by planet b are $\sim 4$ times longer than those of planet c ($\simeq 28$ min against $\simeq 7$ min). The TTV amplitude is proportional to the period of the perturbed planet and to the mass of the perturbing one \citep{agol2005,holman2005}. Therefore, if the two planets had a similar mass, we would expect the outer one to show stronger TTVs. This indicates, then, that the outer planet is the heaviest.\\
The transit times were shifted to the mean ephemeris, and each transit was normalized using the derived normalization coefficients. To model the TTVs, we performed dynamical simulations with the \texttt{mercury} code, version 6.2 \citep{chambers1999}. The integrations were executed with a Bulirsch-Stoerwe algorithm. For each of them, we identified the transit times by interpolating the passage of the planets through the line of sight. We computed the TTVs by subtracting a linear fit to the transit times. As a compromise between execution time and accuracy of the TTVs with respect to the measured uncertainties, we set the simulations to cover the time span of the Kepler photometry, with a step of 0.4 days, that is, $1/47$th of the lower orbital period.\\
The simulations were included in the Bayesian fit that is described in the next paragraph.

\subsubsection{Bayesian combined fit}\label{modelling}

The Kepler photometry, reduced as described in Sect. \ref{photom} and corrected for the TTVs as explained in Sect. \ref{measttvs}, and the SOPHIE radial velocities were fitted together using the planet analysis and small transit investigation software (\texttt{PASTIS}) described in \cite{diaz2014}. This software has been primarily designed to calculate the Bayesian odds ratios between competing scenarios in planetary validation problems. \texttt{PASTIS} allows simultaneously modeling of several data sets and obtaining samples of the parameter posteriors with a Markov chain Monte Carlo (MCMC) algorithm.\\
An exploration phase was started at random points drawn from the priors listed in Table \ref{priors}. From the chains computed in this phase, we used the one with the highest likelihood for the starting values of the final MCMC set. The solution of the exploration phase with the highest likelihood has the lowest eccentricities.\\
To take into account the differences between the stellar models, we used four evolutionary tracks as input for the stellar parameters: Dartmouth \citep{dotter2008}, PARSEC \citep{bressan2012}, StarEvol \cite[Palacios, {priv. com.};][]{Lagarde2012}, and Geneva \citep{mowlavi2012}. However, the intrinsic uncertainties in the models were not taken into account. We ran twenty-five chains of about $10^5$ steps for each of the stellar evolutionary tracks. At each step of the MCMC, the model light curves were oversampled and then binned by a factor 10, to correct for the distortions in the signal due to the finite integration time \citep{kipping2010}. We derived the stellar density $\rho_\star$ from the spectroscopic \teff, \logg, and \feh\ (Sect. \ref{host}) and set it, together with the spectroscopic \teff\ and \feh, as a jump parameter, with normal priors for all three of them. For each planet, we used Gaussian priors for the period $P$ and the primary transit epoch $T_0$ and non-informative priors for the argument of periastron $\omega$, the inclination $i$, and the eccentricity $e$. We stress this last point: without imposing zero eccentricities, we consistently measured these key parameters by taking into account all the available information, TTVs included.\\
We used uniform priors for the coefficients of quadratic limb-darkening, for the planetary-to-stellar radius ratio $R_p/R_\star$, and for the radial velocity amplitude $K$. For Kepler-117 b, we fitted the longitude of the ascending node $\Omega$, too, for which we imposed a uniform prior. The $\Omega$ of planet c was fixed at $180^\circ$ because the symmetry of the problem allows freely choosing one of the two $\Omega$s.\\
We expressed the Kepler normalized flux offset, the systemic velocity, and the RV linear drift with uniform priors (separating LC and SC data in the photometry). Finally, we modeled the instrumental and astrophysical systematic sources of error with a jitter term for Kepler, two for the TTVs (one for each planet), and one for SOPHIE. A uniform prior was assigned to all the jitter terms.\\

After they were sampled, every posterior distribution was thinned according to its correlation length. A combined posterior distribution was derived by taking the same number of points from each stellar evolutionary track. This combined distribution gave the most probable values and the confidence intervals for the system parameters.\\ 
Finally, the derived \logg\ and the posterior stellar radius $R_\star$, \teff, and \feh, together with the magnitudes in Table \ref{tabmag}, were set as priors for another MCMC run to derive the distance of the system using the spectral energy distribution (SED). The magnitudes were fitted to sample the posterior distributions of the distance of the system, the interstellar extinction $E(B-V)$, and the jitter of the SED. The model SED was interpolated from the PHOENIX/BT-Settl synthetic spectral library \citep{allard2012}, scaled with the distance, the stellar radius, and the reddening $E(B-V)$, expressed through a \citet{fitzpatrick1999}  extinction law. For both the distance and the reddening, non-informative priors were used.\\ 

In Table \ref{posteriors} we present the mode and the 68.3\% equal-tailed confidence intervals of the system parameters. According to our analysis, Kepler-117 A is a $\simeq 5$ Gyr old F8V star with two planets in low-eccentricity orbits ($0.0493 \pm 0.0062$ and $0.0323 \pm 0.0033$ for planet b and c), which differ widely in their mass, but less so in their radii: $0.094 \pm 0.033$ \MJ, $0.719 \pm 0.024$ \RJ\ for planet b and $1.84 \pm 0.18$ \MJ, $1.101 \pm 0.035$ \RJ\ for planet c. The planetary radii, in particular, agree with the estimate of \cite{rowe2014}: $\simeq 0.72 \pm 0.14$ \RJ\ for planet b and $\simeq 1.04 \pm 0.20$ \RJ\ for planet c.\\
We remark that the measured drift of the RVs is compatible with 0 \kms; a non-zero drift would have been an indication of a possible third companion in the system that affected the amplitude of the TTVs.

\begin{figure}[!htbp]
\centering
\includegraphics[scale = 0.45]{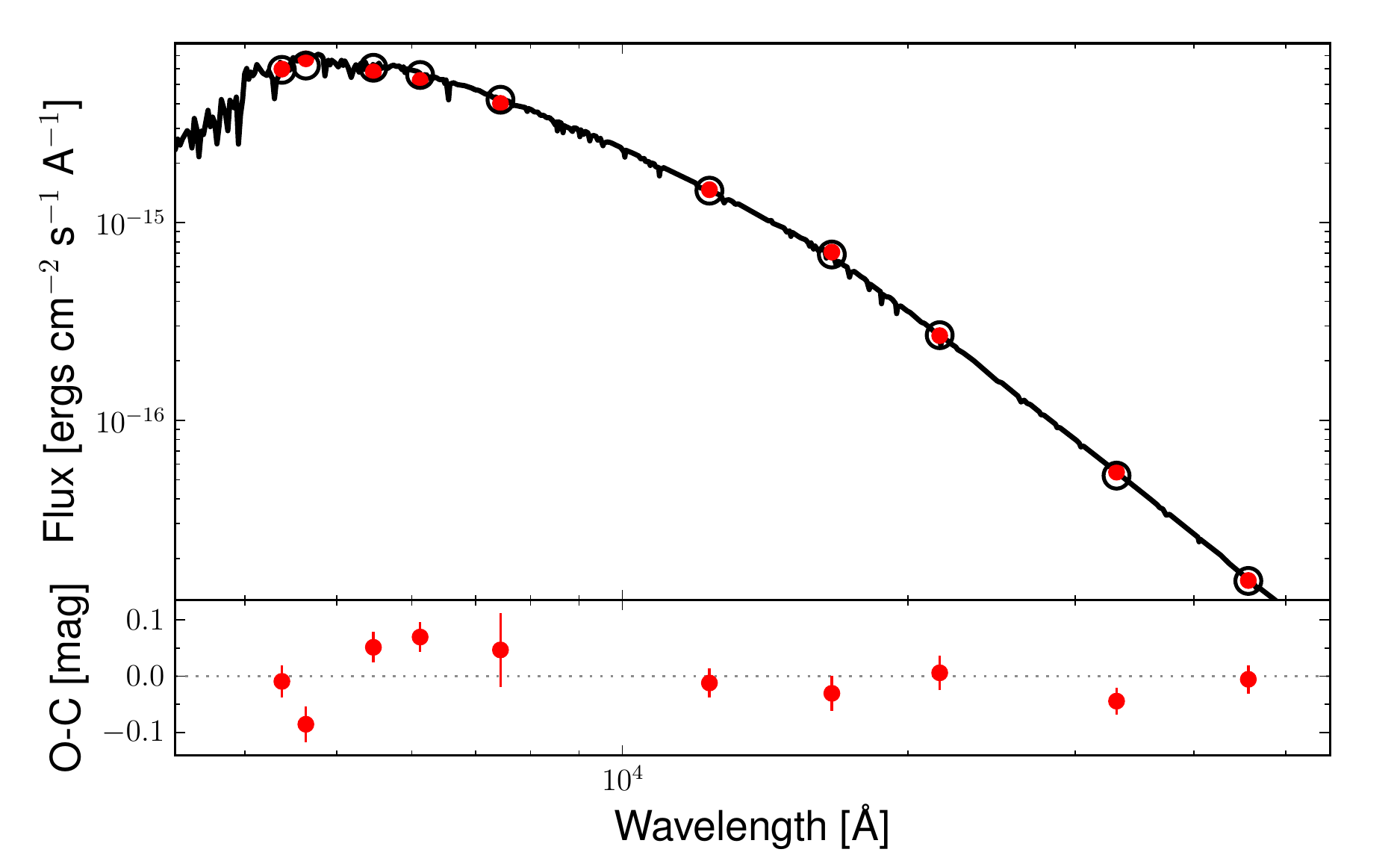}
\caption{Model SED on the photometric bands. The residuals are shown in the lower panel.}
\label{sed}
\end{figure}

\begin{table*}[htbp]
\centering
\caption{\label{posteriors} Planetary and stellar parameters with their 68.3\% central confidence intervals.}
\begin{tabular}{lll}
\hline\hline
\multicolumn{1}{l}{\emph{Stellar parameters from the combined analysis}} & & \smallskip\\
Stellar density $\rho_{\star}$ [$\rho_\odot$]    & 0.291$^{+0.010}_{-0.018}$ & \\
Stellar mass [\Msun]                          & 1.129$^{+0.13}_{-0.023}$  & \\
Stellar radius [\Rsun]                        & 1.606 $\pm$ 0.049      & \\
\teff [K]                                     & 6150 $\pm$ 110         & \\
Metallicity [Fe/H] [dex]                      & -0.04 $\pm$ 0.10       & \\
Derived \logg [cgs]                           & 4.102 $\pm$ 0.019      & \\
Age $t$ [Gyr]                                 & 5.3 $\pm$ 1.4          & \\
Distance of the system [pc]                   & 1430 $\pm$ 50          & \\
Interstellar extinction $E(B - V)$            & 0.057 $\pm$ 0.029      & \\
Quadratic limb-darkening coefficient $u_a$    & 0.382 $\pm$ 0.031      & \\
Quadratic limb-darkening coefficient $u_b$    & 0.187 $\pm$ 0.056      & \\
Stellar RV linear drift [\ms yr$^{-1}$]        & 12 $\pm$ 22            & \\
Systemic velocity (BJD 2456355), $V_{r}$ [\kms]& -12.951 $\pm$ 0.013    & \smallskip\\

\multicolumn{1}{l}{} & \emph{Kepler-117 b} & \emph{Kepler-117 c} \smallskip\\
Orbital period $P$ [days]                   & 18.7959228 $\pm 7.5\e{-6}$ & 50.790391 $\pm 1.4\e{-5}$ \\
Orbital semi-major axis $a$ [AU]            & 0.1445$^{+0.0047}_{-0.0014}$   & 0.2804$^{+0.0092}_{-0.0028}$ \\
Primary transit epoch $T_{0}$ [BJD-2450000] & 4978.82204 $\pm 3.5\e{-4}$  & 4968.63205 $\pm 2.5\e{-4}$ \\
Orbital eccentricity $e$                    & 0.0493 $\pm$ 0.0062        & 0.0323 $\pm$ 0.0033 \\
Argument of periastron $\omega$ [deg]       & 254.3 $\pm$ 4.1            & 305.0 $\pm$ 7.5 \\
Orbital inclination $i$ [deg]               & 88.74 $\pm$ 0.12           & 89.64 $\pm$ 0.10$^{\dagger}$ \\
Transit impact parameter $b_{prim}$          & 0.446 $\pm$ 0.032          & 0.268$^{+0.036}_{-0.087}$$^{\ddagger}$ \\
Transit duration $T_{14}$ [h]                & 7.258 $\pm$ 0.020          & 10.883 $\pm$ 0.031 \\
Scaled semi-major axis $a/R_\star$           & 19.67 $\pm$ 0.37           & 38.18 $\pm$ 0.72 \\
Radius ratio $R_p/R_\star$                   & 0.04630 $\pm$ 0.00025      & 0.07052 $\pm$ 0.00034 \\
Radial velocity semi-amplitude $K$ [\ms]    & 6.5 $\pm$ 2.1              & 90.4 $\pm$ 7.0 \\
Longitude of the ascending node $\Omega$ [deg] & 177.9 $\pm$ 5.6         & 180 (fixed) \smallskip\\

Planet mass $M_{p}$ [M$_J$]                   & 0.094 $\pm$ 0.033         & 1.84 $\pm$ 0.18 \\
Planet radius $R_{p}$[R$_J$]                  & 0.719 $\pm$ 0.024         & 1.101 $\pm$ 0.035 \\
Planet density $\rho_p$ [$g\;cm^{-3}$]        & 0.30 $\pm$ 0.11           & 1.74 $\pm$ 0.18 \\
Planet surface gravity, $\log$\,$g_{p}$ [cgs] & 2.67$^{+0.10}_{-0.17}$       & 3.574 $\pm$ 0.041 \\
Planet equilibrium temperature, $T_{eq}$ [K]$^{\ast}$ & 984 $\pm$ 18       & 704 $\pm$ 15 \smallskip\\

\multicolumn{3}{l}{\emph{Data-related parameters}} \smallskip\\
Kepler jitter (LC)  [ppm] & 67$^{+16}_{-39}$  & \\
Kepler jitter (SC)  [ppm] & 0$^{+12}$        & \\
SOPHIE jitter$^a$[\ms]        & 0$^{+25}$         & \\ 
SED jitter [mags]          & 0.043$^{+0.026}_{-0.013}$ & \\
TTV1 jitter [min]          & 0.95 $\pm$ 0.48  & \\
TTV2 jitter [min]          & 0.90 $\pm$ 0.62  & \\
\\

\hline
\end{tabular}
\begin{list}{}{}
\item $^{\dagger}$ From the posterior distribution, reflected with respect to $i = 90^\circ$.
\item $^{\ddagger}$ Reflected as the inclination, with respect to $b = 0$.
\item $^{\ast}$ $T_{eq}=T_{\mathrm{eff}}\left(1-A\right)^{1/4}\sqrt{\frac{R_\star}{2 a}}$, with $A \, \mathrm{(planet \, albedo)}\; =0$.
\item $^a$ This value is compatible with the low level of activity observed in the photometry (Sect. \ref{activity}) and confirms the estimate of the error bars on the RVs (Sect. \ref{spectro}). 
\item \Msun $= 1.98842\e{30}$~kg, \Rsun $= 6.95508\e{8}$~m, M$_J = 1.89852\e{27}$~kg, R$_J$ = 71492000~m
\end{list}
\end{table*} 

\section{Discussion}\label{disc}

\subsection{Modulation of the TTVs}\label{modu}

\begin{figure*}[htbp]
\centering
\includegraphics[scale = 0.45]{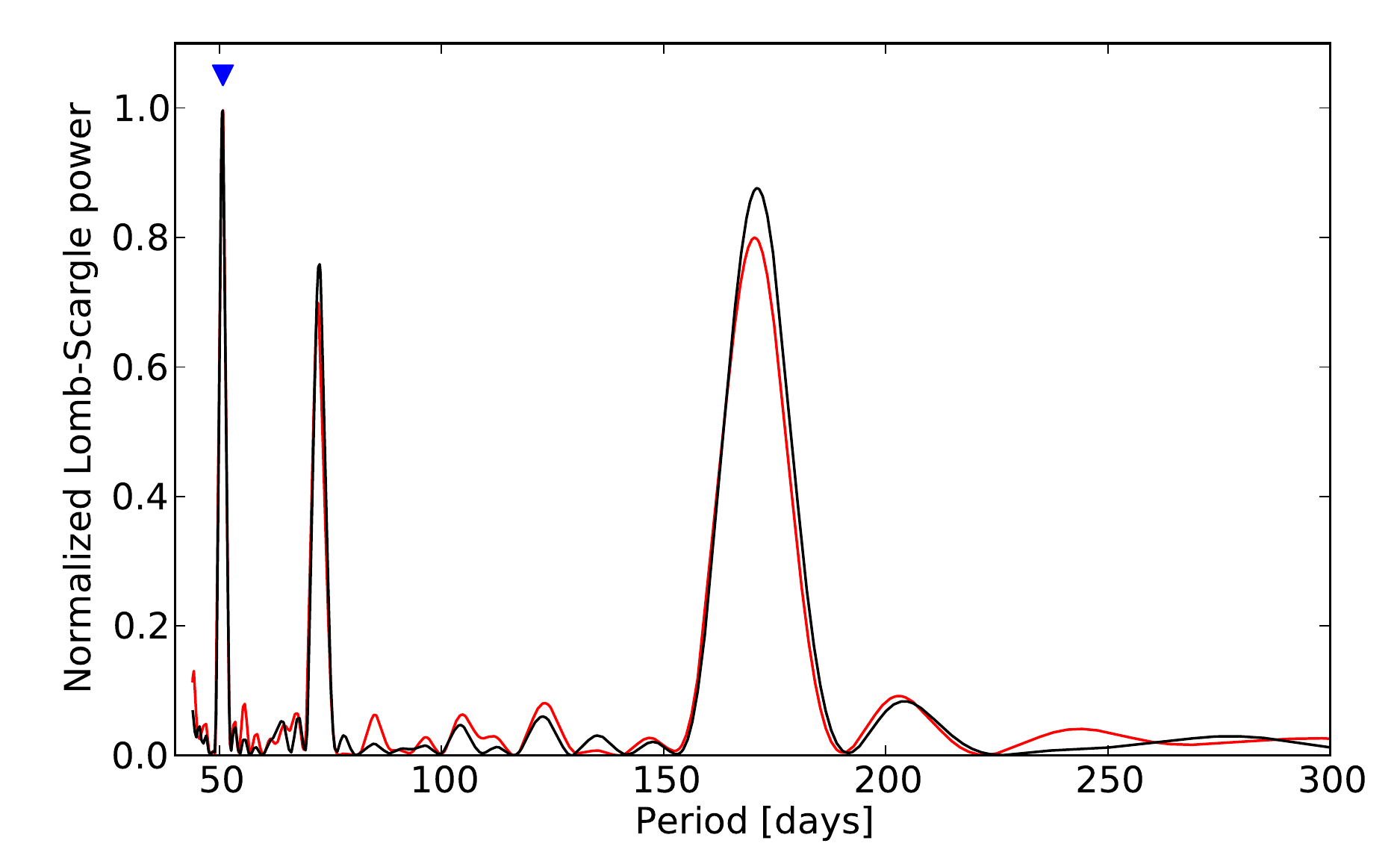}
\includegraphics[scale = 0.45]{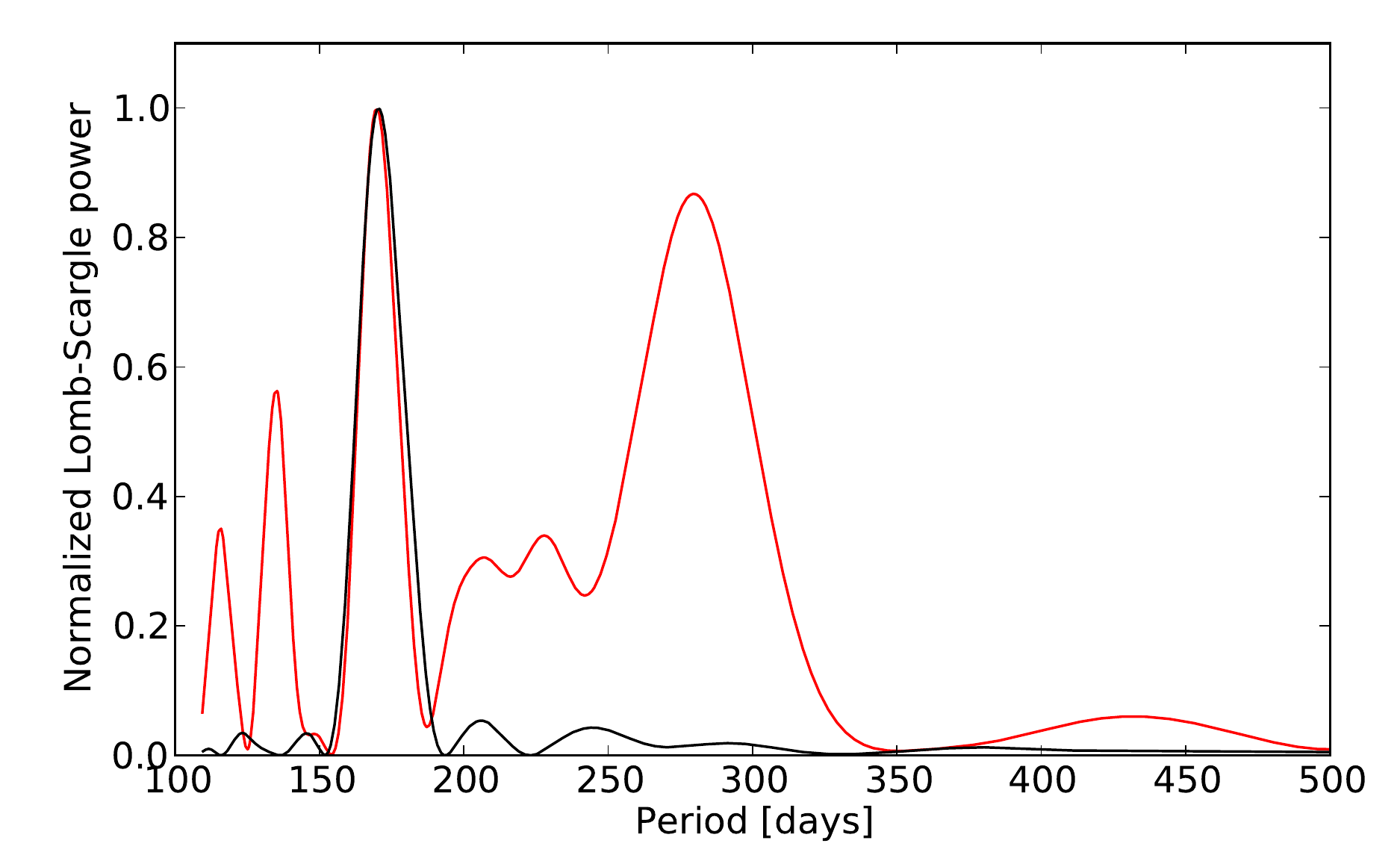}
\caption{Lomb-Scargle periodogram of the measured (red) and modeled (black) TTVs for planet b (left) and c (right). The blue triangle in the plot on the left indicates the orbital period of planet c.}
\label{lsttv}
\end{figure*}

We tested the robustness of our result by inspecting the periodic modulation of the measured and the modeled TTVs. To do this, we compared their Lomb-Scargle periodograms (Fig. \ref{lsttv}). The main peaks coincide for both planets and also agree with the periodicities found by  \citealp{mazeh2013} \citep[see also][]{ofirtoulouse}. The periodogram of the modeled TTVs of planet b reproduces that of the measured TTVs well. Some of the peaks of planet c, on the other hand, are due to noise. This can be explained by the different amplitude of the signal in the two cases.

\subsection{Constraining the system parameters with the TTVs}\label{constr}

\begin{figure*}[!htbp]
\centering
\includegraphics[scale = 0.61]{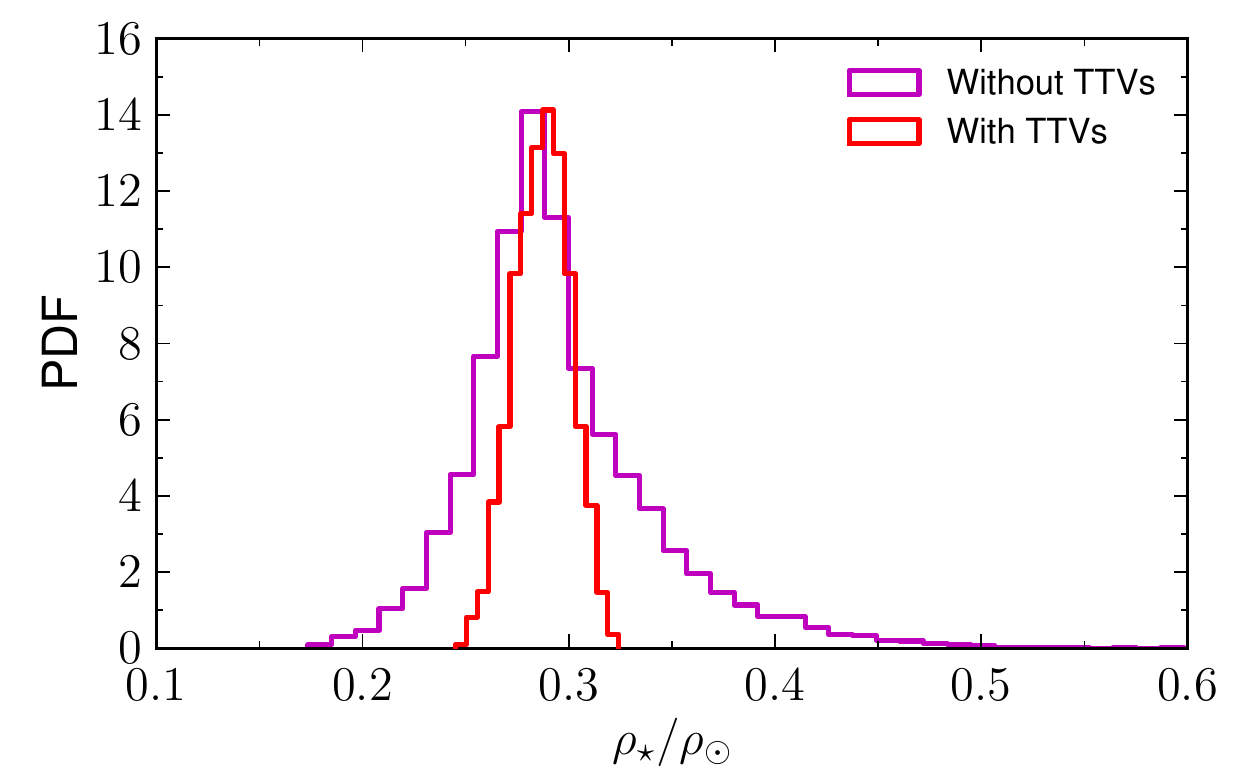}
\includegraphics[scale = 0.61]{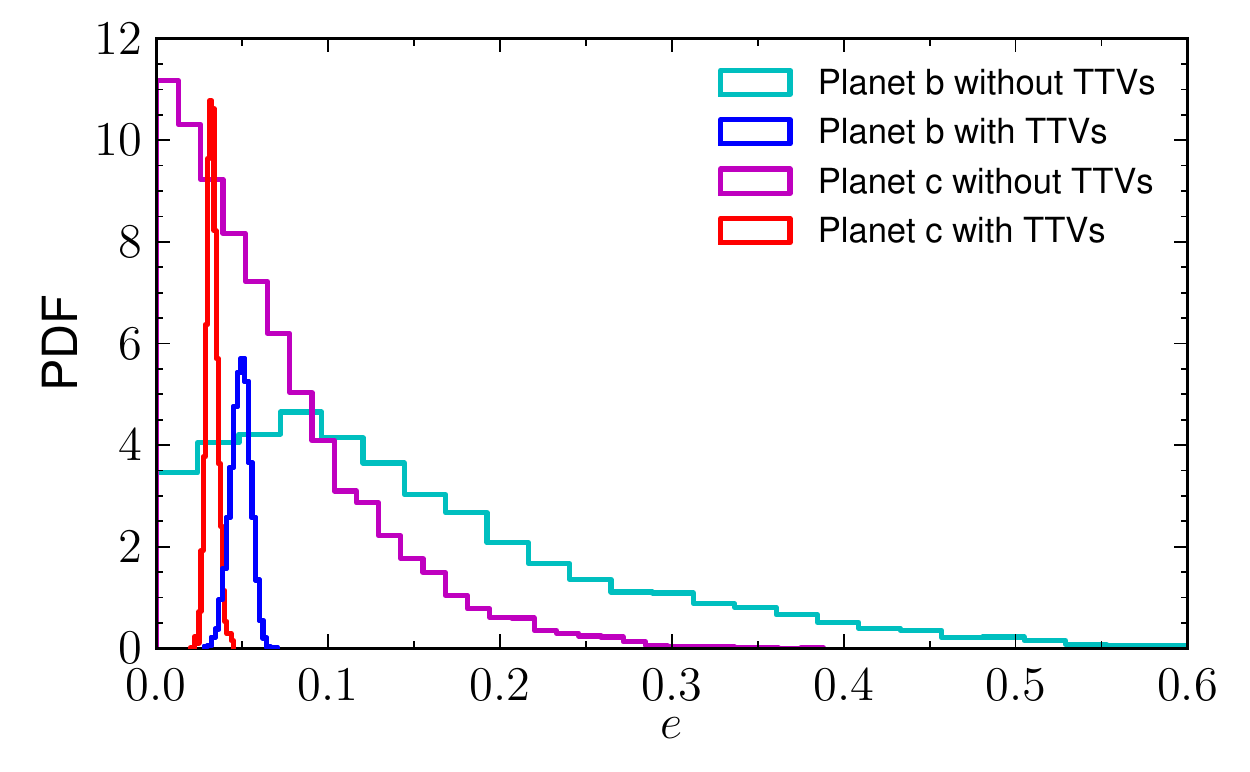}
\includegraphics[scale = 0.61]{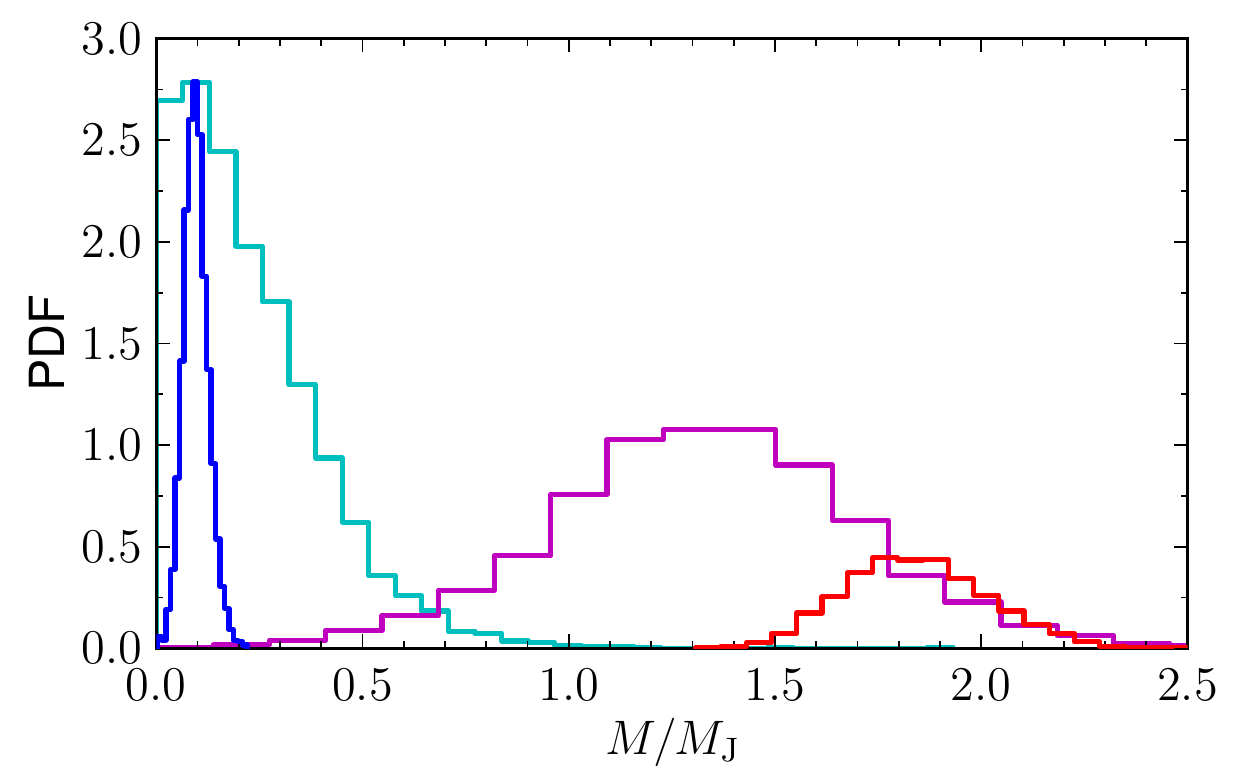}
\includegraphics[scale = 0.61]{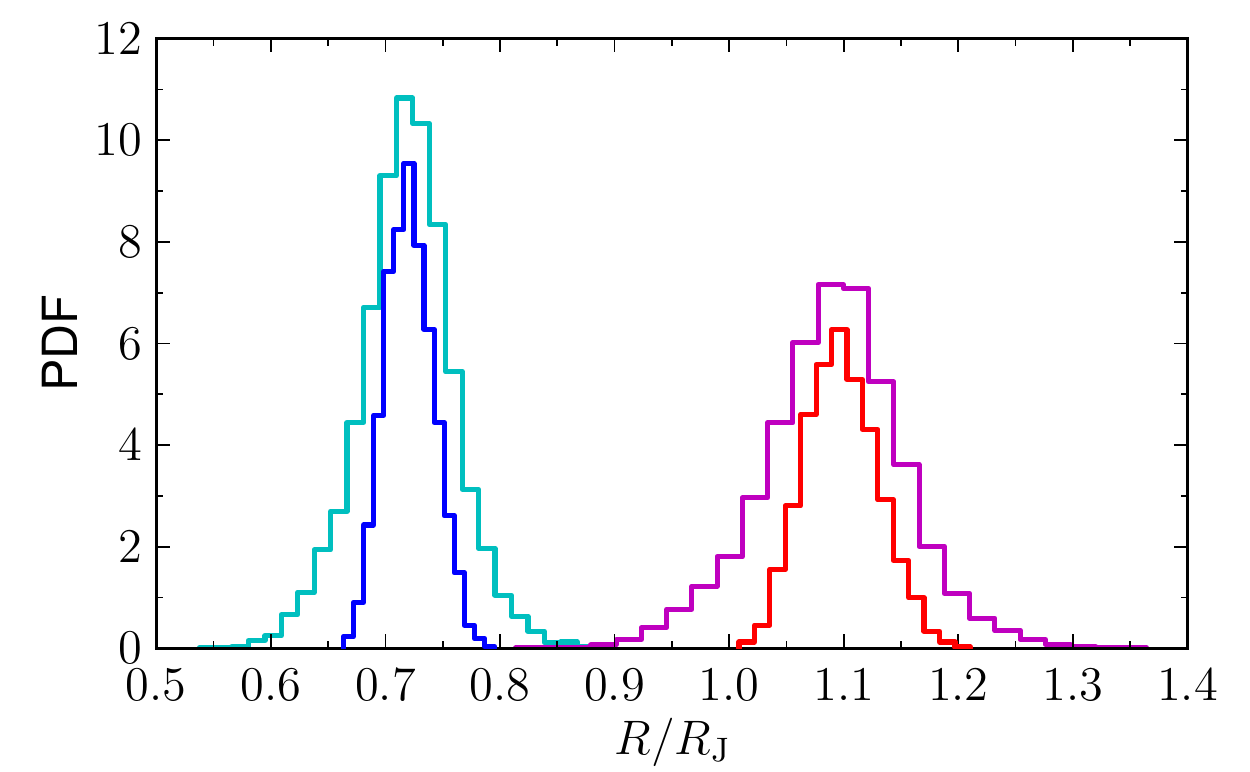}
\includegraphics[scale = 0.6]{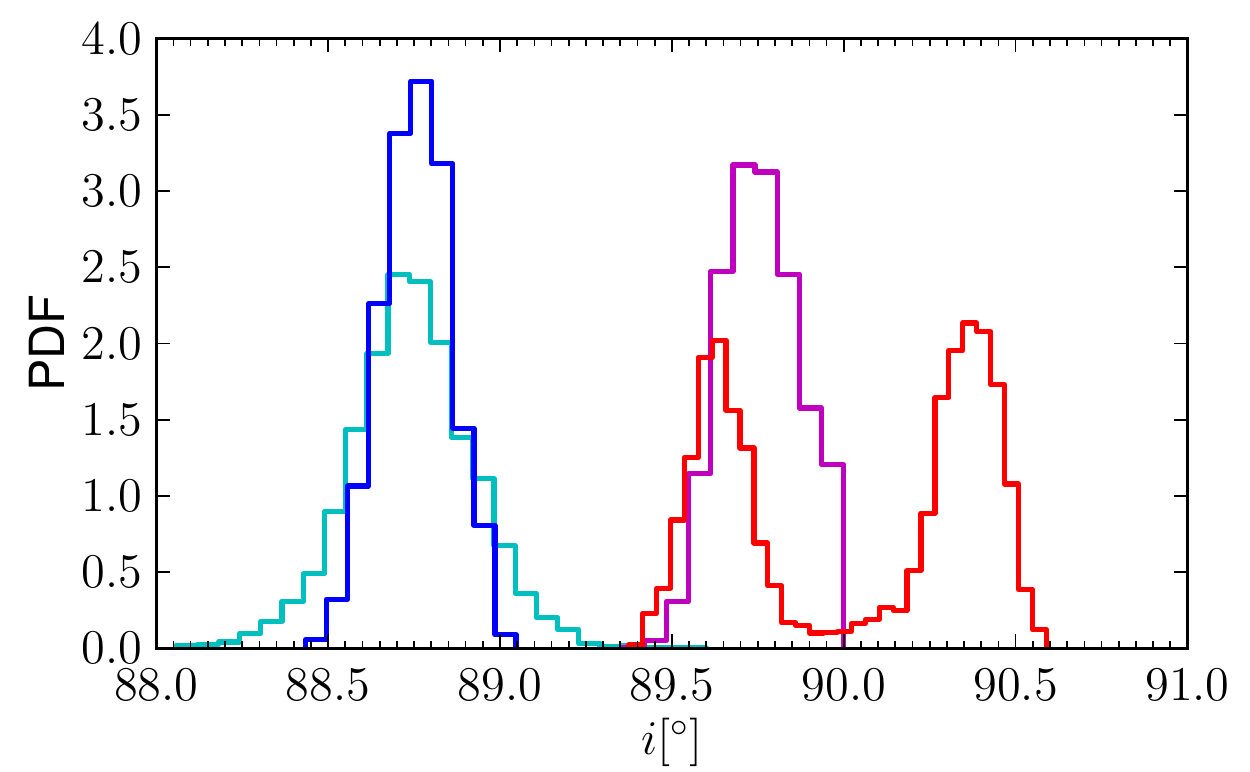}
\caption{\textit{Upper left:} probability density function (PDF) of the stellar density including or excluding the TTVs; to make the plot more readable, the PDF using the TTVs is divided by 2. The prior from spectroscopy (HIRES, Dartmouth) is shown for comparison. \textit{Upper right:} the planetary eccentricities from different sets of data; the PDF using the TTVs is divided by 12. The color code is the same in the following plots. \textit{Central line:} planetary masses and radii from the fit with and without TTVs. The PDF of the masses using the TTVs is divided by 5, that of the radii by 2. \textit{Bottom:} orbital inclinations from different sets of data.}
\label{comparisons}
\end{figure*}

\begin{figure}[!htbp]
\centering
\includegraphics[scale = 0.45]{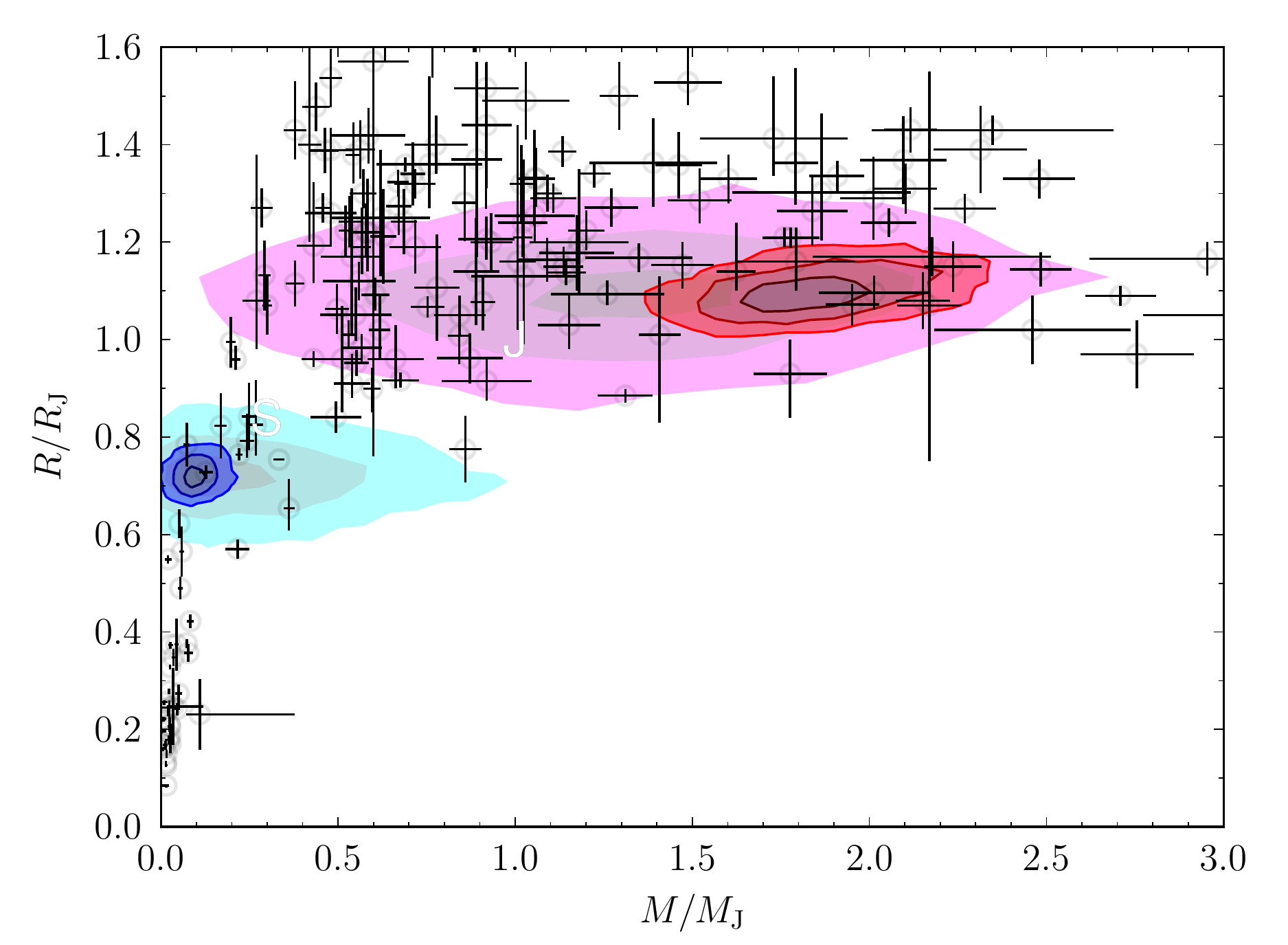}
\caption{Mass-radius diagram for the solutions without (cyan for planet b, magenta for planet c) and with (blue for planet b, red for planet c) TTVs. The blue and red solutions are those indicated in Table \ref{posteriors}. The colors, from the center to the edge of the regions, correspond to the 39.3\%, 86.5\%, and 98.9\% joint confidence intervals. Jupiter and Saturn (labeled J and S) are marked for comparison. The other planet parameters were taken from \cite{wright2011}.}
\label{mtor}
\end{figure}

The RV amplitude produced by Kepler-117 b ($6.5 \pm 2.1$ \ms) is close to the sensitivity of SOPHIE for a $\simeq 14.5$ mag$_V$ star. Indeed, the SOPHIE RVs alone do not have the precision required to measure this planet's mass and can only provide an upper limit. Including the TTVs allowed us to accurately determine this parameter.\\
The combined fit of the TTVs reduces the uncertainties on other parameters, too. This is because the amplitude of the TTVs is mainly determined by orbital separations, periods, and eccentricities of the orbits and masses ratios of the planets \citep{agol2005}. The strong constraint on the eccentricities, combined with the constraint on the stellar density (which is determined with a precision similar to the precisions achieved in asteroseismology), reduces the error bars on the planetary parameters. Once again, we stress that the derived uncertainties do not include the uncertainties on all the models of stellar atmospheres and evolutionary tracks we used.\\
To check the impact of the TTVs on the combined fit, we ran \texttt{PASTIS} without them. The different posteriors of the most affected parameters, with or without the TTVs, are compared in Fig. \ref{comparisons} and \ref{mtor}. The mass of planet b presents the most evident difference because of its poorly constrained RV amplitude: its value reaches from an upper limit (0.28 \MJ\ at 68.3\% confidence level) to a better constrained value ($0.094 \pm 0.033$ \MJ). The difference is smaller for planet c because the amplitude of the RVs is larger and better fitted. However, its uncertainty is roughly reduced by 40\%. This indicates that, if possible, including the TTVs in the combined fit is more effective than fitting them \textit{a posteriori}, using a set of orbital parameters derived without considering them.\\
We remark that the mass of planet c found with the RVs alone and with the TTVs are fully compatible. Therefore, the observed TTVs are completely explained by the two planets, within the data error bars. This agrees with the absence of any RV drift (Sect. \ref{modelling}).\\

The transit signature is degenerate with respect to the stellar hemisphere the planet covers, while the TTVs are not. In a two-planet scenario, this can lead to strong correlations between the two inclinations. While running \texttt{PASTIS}, we therefore constrained one of the transits in one of the hemispheres and left the other free to vary. As the inclination of planet b is lower than that of planet c, the inclination of planet b was limited to one hemisphere ($50^\circ < i < 90^\circ$) and that of planet c was left free to vary between both ($89^\circ < i < 91^\circ$). In spite of this, our fit allowed both hemispheres to be transited by planet c because the final inclination of its orbit is almost symmetric with respect to the stellar equator (Fig. \ref{comparisons}, bottom line).  The solutions with $i > 90^{\circ}$ are compatible with those without TTVs (all $< 90^\circ$) at $1\sigma$. In particular, for planet c, we found $89.64 \pm 0.10^\circ$ with TTVs and $i = 89.75 \pm 0.13^\circ$ without them.\\
Using the stellar inclination (Sect. \ref{activity}) and the system parameters, we calculated the expected amplitude of the Rossiter-McLaughlin effects following Eq. 11 of \cite{gaudi2007}. For planet b, we found $10.9 \pm 3.0$ \ms, for planet c $79 \pm 13$ \ms. Measuring the spin-orbit misalignment would then be possible for planet c, but the transit duration ($\simeq 11$ hours) would require a joint effort from different locations to cover a whole transit.\\
The difference between the resulting longitude of the ascending node $\Omega$ for planet b ($177.9 \pm 5.6^\circ$) and that of planet c (fixed to $180^\circ$) is compatible with $0^\circ$. Combined with the similar inclinations, this implies two almost coplanar orbits. As most of the Kepler planetary systems \citep{fabrycky2014}, Kepler-117 clearly has a flat configuration of the orbits.\\

We ran an MCMC set without the RVs to determine the reliability of the fit. As expected, the posterior distributions are the same as with the RVs. Systems with low-mass planets presenting TTVs, which are challenging for the RV observations, would benefit from the approach used in this paper.\\

\begin{figure*}[!htbp]
\centering
\includegraphics[scale = 0.6]{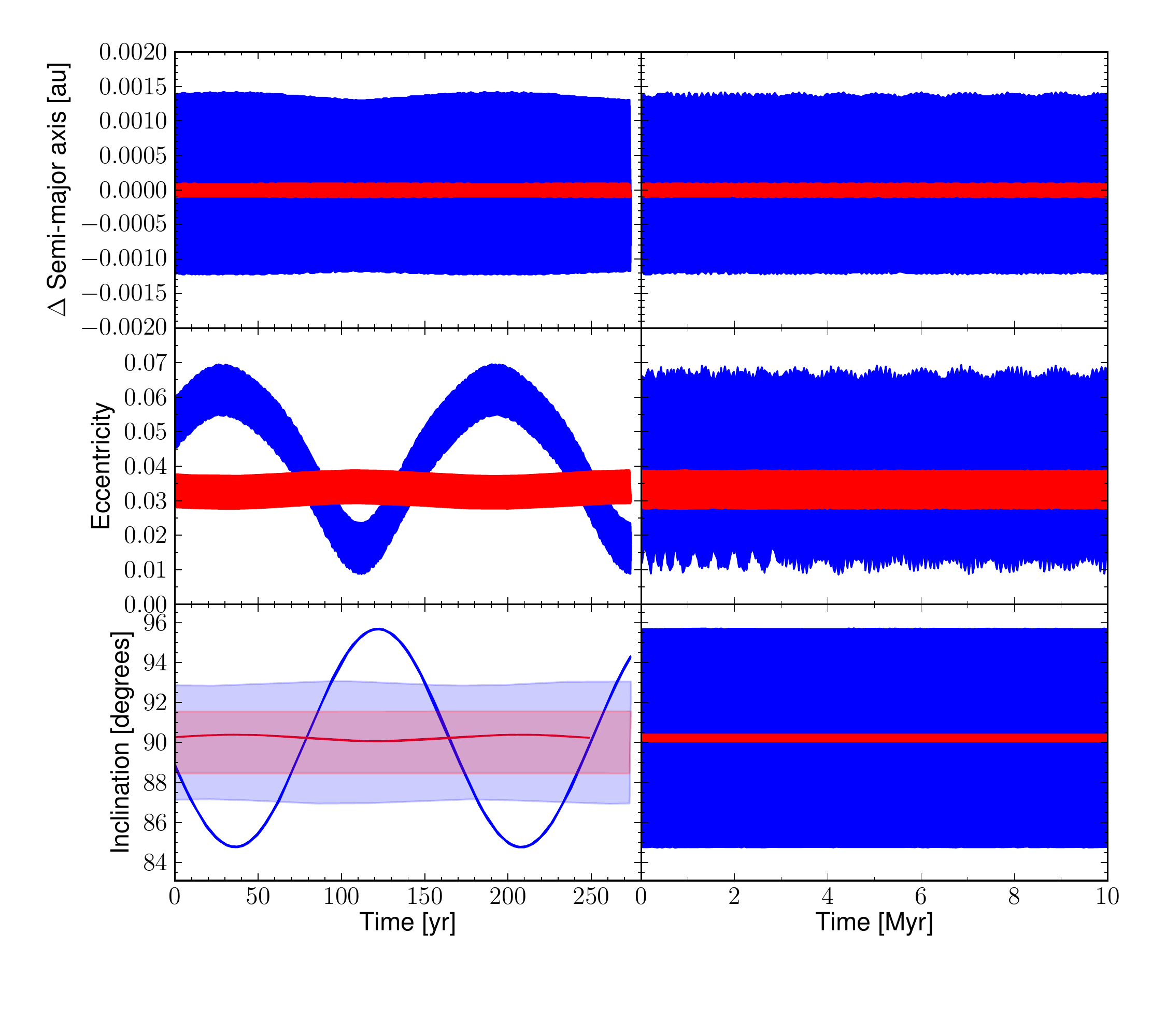}
\caption{Evolution of semi-major axes (top), eccentricities (center), and orbital inclinations (bottom) over a 10 Myr simulation of the most probable solution. The respective mean has been subtracted from the two semi-major axes. On the left, a zoom on the first $300$ yr; on the right, the variation intervals of the parameters. In blue planet b, in red planet c. The  shaded regions in the left panel of the inclinations correspond to the values resulting in a transit (see Eq. 7 of \citealp{winn2010}).}
\label{stab}
\end{figure*}

Finally, we verified that the configuration of the most probable solution is dynamically stable. We ran \texttt{mercury} over a time span of 10 Myr (Fig. \ref{stab}). The semi-major axes, eccentricities, and orbital inclinations oscillate over a time scale of around 200 years, but all the parameters are stable in the long term.

\subsection{A third non-transiting companion around Kepler-117?}

\begin{figure}[htbp]
\centering
\includegraphics[scale = 0.4]{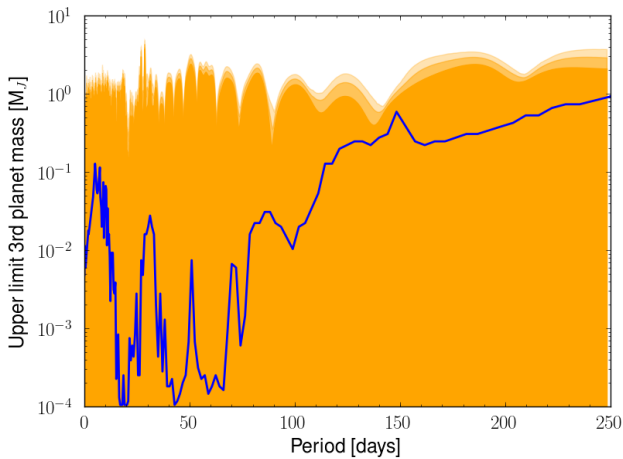}
\caption{Maximum mass of a third possible companion as a function of the orbital period, based on the RV (orange filled areas) and TTV (blue line) observations. The color, from darker to lighter, corresponds to the 68.27\%, 95.45\%, and 99.73\% confidence intervals from the RV data. The blue line represents a $3\sigma$ difference in the residuals on the TTVs from the best solution with two planets.}
\label{3rdlim}
\end{figure}

The possibility of a third non-transiting companion can be probed with the RVs and the TTVs. As already mentioned in Sect. \ref{modelling}, the absence of stellar drift in the RVs brings no evidence of the possible presence of a third non-transiting planet in the system. Moreover, the agreement between the mass of planet c, found with the RVs and with or without the TTVs (Sect. \ref{constr}), shows that the TTVs are not affected by a non-transiting body.\\
A more precise constraint can be obtained by subtracting the modeled RVs of the two planets from the RV measurements. We folded the residuals for several periods and fitted them with a sinusoid. The amplitude of the sinusoid and the mass of the star (Table \ref{posteriors}) allow extracting the maximum mass of the possible companion.\\
The result is plotted in Fig. \ref{3rdlim} (filled regions) for the 68.27\%, 95.45\%, and 99.73\% confidence intervals. The RVs allow the presence of a Jupiter-mass planet for some orbital periods. The TTVs, however, impose a stronger constraint, since including a third body with the combination of mass and period allowed by the RVs (with the simplifying assumptions of a circular orbit and $90^\circ$ inclination) would not fit the TTVs. The black line in the plot represents a $3\sigma$ difference in the residuals between the fit of the TTVs with a third planet and the best solution with two.\\
Therefore, under some simplifying assumptions, the presence of a non-detected third companion above $\sim 0.1$ \MJ\ on an orbit shorter than $\sim 100$ days, as well as that of a giant companion with an orbit shorter than $\sim 250$ days, is very unlikely.\\

\section{Summary and conclusions}\label{conclusions}

We presented the combined analysis of the Kepler photometry, the TTVs, the SOPHIE RVs, and the spectroscopic observations of Kepler-117. This allowed us to measure the stellar, planetary, and orbital parameters of the system. According to our analysis, Kepler-117 A is an F8-type main-sequence star that is about 5 Gyr old. The system is composed of two planets. Kepler-117 c, the outer one, has a period of $\simeq 51$ days and a $\sim 2$ \MJ\ mass. Kepler-117 b, the inner one, has a period of $\simeq 19$ days and a $\sim 30 \, M_\mathrm{Earth}$ mass. The latter produces a RV semi-amplitude of $6.5 \pm 2.1$ \ms, close to the limit of sensitivity of SOPHIE for faint magnitudes. Therefore, its mass and eccentricity cannot be obtained from the RVs alone, even though the derived upper limits confirm that the transiting body belongs to the planet realm.\\
Our analysis shows that the inclusion of the TTVs in the combined fit allows tightly constraining the mass of the lighter planet. Taking into account the TTVs in the fit also allows a better determination of the other system parameters. The stellar density was tightly constrained despite a loose prior on the spectroscopic $\log g$. The planetary radii were strongly constrained, as were the orbital eccentricities, even if small (around $\sim 0.03-0.05$ for both orbits). Measuring the eccentricity accurately is important for testing the dynamical models of young systems with giant planets. Simulations show that the complex evolution of systems with two planets can be the result of ejection or merging in systems with three planets and can lead to stable, resonant, and low-eccentricity orbits \citep[e.g.,][]{lega2013}.\\
While the RV- and TTV-measured mass agree for planet c, the same comparison for planet b would benefit from a spectrograph with higher sensitivity than SOPHIE.  Only a small part of the planets known to date have their mass measured with both RVs and TTVs \citep[e.g.,][]{barros2014}. As observed by \cite{weiss2014}, planets smaller that four Earth radii with a mass measured with TTVs are systematically lower in the mass-radius diagram than those discovered by RV surveys. This could be due to non-detected companions that might dampen the TTVs, causing a systematic underestimation of the masses, or to a lower density of the planets that show TTVs. Indeed, in the compact multiplanetary systems that are likely to produce observable TTVs, planets with lower masses for a given size are more likely to reach stable orbits \citep{jontof-hutter2014}. In this context, it is remarkable that the RV- and the TTV-measured mass of Kepler-117 c agree. In addition, under some simplifying assumptions, the TTVs almost exclude a non-detected $\sim 0.1$ \MJ\ companion with an orbit shorter than $\sim100$ days, as well as a giant companion with an orbit shorter than $\sim 250$ days.\\
We cannot exclude that while in this particular case the conditions are fulfilled for the combined fit to be effective, this is not the case in general. In fact, TTVs with sufficiently high amplitude are necessary. This system exhibits significant TTVs even though the orbital period ratio of the planets is far from an exact low-order mean motion resonance, for which strong TTVs are expected \citep[e.g.,][]{lithwick2012}. The orbital period ratio between planet c and b ($\simeq 2.7$) places this system on the wide side of the 5:2 mean motion resonance. The overabundance of systems with period ratios some percent higher than resonant values than those with a ratio slightly lower than these ones has been well established for the Kepler systems in the case of first-order resonances, that is, 2:1 or 3:2 \citep[see][e.g.]{lissauer2011}. The explanation is given in terms of tidal dissipation related to disk-planet or star-planet interactions for close-in orbits \citep{lithwick2012,batygin2013,delisle2014}, causing the orbital periods to diverge. 
To date, this piling-up appears only for some first-order resonances in the Kepler systems. Instead, the rest of the period-ratio distribution, including higher order resonances, remains flat \citep[e.g.,][]{batygin2013,fabrycky2014}.\\ 
We verified that a system with the most probable solution of our analysis is dynamically stable. However, we noted the eccentricities and the inclinations show small oscillations, that do not affect the stability of the system. We found the planetary orbits to be almost coplanar. This places Kepler-117 in the most common population of the Kepler multiplanetary systems with a flat configuration, as highlighted by \cite{fabrycky2014}.\\
In conclusion, a deeper understanding of the dynamics of orbital resonances is needed to better reconstruct the history of Kepler-117, which adds valuable information to our knowledge of multiplanetary systems.
  
\begin{acknowledgements}
This paper includes data collected by the Kepler mission. Funding for the Kepler mission is provided by the NASA Science Mission directorate. We made use of the Mikulski Archive for Space Telescopes (MAST). Support for MAST for non-HST data is provided by the NASA Office of Space Science via grant NNX09AF08G and by other grants and contracts.\\
We thank the technical team at the Observatoire de Haute-Provence for their support with the SOPHIE instrument and the 1.93 m telescope and in particular for the essential work of the night assistants. Financial support for the SOPHIE observations comes from the Programme National de Planetologie (PNP) of CNRS/INSU, France is gratefully acknowledged. We also acknowledge support from the French National Research Agency (ANR-08- JCJC-0102-01).\\
The team at LAM acknowledges support by CNES grants 98761 (SCCB), 426808 (CD), and 251091 (JMA). AS acknowledge the support from the European Research Council/European Community under the FP7 through Starting Grant agreement number 239953. A.S. is supported by the European Union under a Marie Curie Intra-European Fellowship for Career Development with reference FP7-PEOPLE-2013-IEF, number 627202. ASB acknowledges funding from the European Union Seventh Framework Programme (FP7/2007-2013) under Grant agreement No. 313014 (ETAEARTH).\\
We thank John Chambers for his explanations about the use of \texttt{mercury} and Rosemary Mardling for the fruitful discussions about the dynamic of three-body systems.\\
This research was made possible through the use of data from different surveys: the AAVSO Photometric All-Sky Survey (APASS), funded by the Robert Martin Ayers Sciences Fund; the Two Micron All Sky Survey, which is a joint project of the University of Massachusetts and the Infrared Processing and Analysis Center/California Institute of Technology, funded by the National Aeronautics and Space Administration and the National Science Foundation; the Wide-field Infrared Survey Explorer, which is a joint project of the University of California, Los Angeles, and the Jet Propulsion Laboratory/California Institute of Technology, funded by the National Aeronautics and Space Administration.\\
This research has made reference to the Exoplanet Orbit Database and the Exoplanet Data Explorer at exoplanets.org.\\

\end{acknowledgements}

\bibliographystyle{aa}
\bibliography{KOI209.bib}

\begin{thebibliography}{57}
\expandafter\ifx\csname natexlab\endcsname\relax\def\natexlab#1{#1}\fi

\bibitem[{{Agol} {et~al.}(2005){Agol}, {Steffen}, {Sari}, \&
  {Clarkson}}]{agol2005}
{Agol}, E., {Steffen}, J., {Sari}, R., \& {Clarkson}, W. 2005, \mnras, 359, 567

\bibitem[{{Allard} {et~al.}(2012){Allard}, {Homeier}, \&
  {Freytag}}]{allard2012}
{Allard}, F., {Homeier}, D., \& {Freytag}, B. 2012, Royal Society of London
  Philosophical Transactions Series A, 370, 2765

\bibitem[{{Baranne} {et~al.}(1996){Baranne}, {Queloz}, {Mayor}, {Adrianzyk},
  {Knispel}, {Kohler}, {Lacroix}, {Meunier}, {Rimbaud}, \& {Vin}}]{baranne1996}
{Baranne}, A., {Queloz}, D., {Mayor}, M., {et~al.} 1996, \aaps, 119, 373

\bibitem[{{Barros} {et~al.}(2013){Barros}, {Bou{\'e}}, {Gibson}, {Pollacco},
  {Santerne}, {Keenan}, {Skillen}, \& {Street}}]{barros2013}
{Barros}, S.~C.~C., {Bou{\'e}}, G., {Gibson}, N.~P., {et~al.} 2013, \mnras,
  430, 3032

\bibitem[{{Barros} {et~al.}(2014){Barros}, {D{\'{\i}}az}, {Santerne}, {Bruno},
  {Deleuil}, {Almenara}, {Bonomo}, {Bouchy}, {Damiani}, {H{\'e}brard},
  {Montagnier}, \& {Moutou}}]{barros2014}
{Barros}, S.~C.~C., {D{\'{\i}}az}, R.~F., {Santerne}, A., {et~al.} 2014, \aap,
  561, L1

\bibitem[{{Barros} {et~al.}(2011){Barros}, {Pollacco}, {Gibson}, {Howarth},
  {Keenan}, {Simpson}, {Skillen}, \& {Steele}}]{barros2011}
{Barros}, S.~C.~C., {Pollacco}, D.~L., {Gibson}, N.~P., {et~al.} 2011, \mnras,
  416, 2593

\bibitem[{{Batygin} \& {Morbidelli}(2013)}]{batygin2013}
{Batygin}, K. \& {Morbidelli}, A. 2013, \aj, 145, 1

\bibitem[{{Boisse} {et~al.}(2010){Boisse}, {Eggenberger}, {Santos}, {Lovis},
  {Bouchy}, {H{\'e}brard}, {Arnold}, {Bonfils}, {Delfosse}, {Desort},
  {D{\'{\i}}az}, {Ehrenreich}, {Forveille}, {Gallenne}, {Lagrange}, {Moutou},
  {Udry}, {Pepe}, {Perrier}, {Perruchot}, {Pont}, {Queloz}, {Santerne},
  {S{\'e}gransan}, \& {Vidal-Madjar}}]{boisse2010}
{Boisse}, I., {Eggenberger}, A., {Santos}, N.~C., {et~al.} 2010, \aap, 523, A88

\bibitem[{{Borucki} {et~al.}(2011){Borucki}, {Koch}, {Basri}, {Batalha},
  {Boss}, {Brown}, {Caldwell}, {Christensen-Dalsgaard}, {Cochran}, {DeVore},
  {Dunham}, {Dupree}, {Gautier}, {Geary}, {Gilliland}, {Gould}, {Howell},
  {Jenkins}, {Kjeldsen}, {Latham}, {Lissauer}, {Marcy}, {Monet}, {Sasselov},
  {Tarter}, {Charbonneau}, {Doyle}, {Ford}, {Fortney}, {Holman}, {Seager},
  {Steffen}, {Welsh}, {Allen}, {Bryson}, {Buchhave}, {Chandrasekaran},
  {Christiansen}, {Ciardi}, {Clarke}, {Dotson}, {Endl}, {Fischer}, {Fressin},
  {Haas}, {Horch}, {Howard}, {Isaacson}, {Kolodziejczak}, {Li}, {MacQueen},
  {Meibom}, {Prsa}, {Quintana}, {Rowe}, {Sherry}, {Tenenbaum}, {Torres},
  {Twicken}, {Van Cleve}, {Walkowicz}, \& {Wu}}]{borucki2011}
{Borucki}, W.~J., {Koch}, D.~G., {Basri}, G., {et~al.} 2011, \apj, 728, 117

\bibitem[{{Bouchy} {et~al.}(2011){Bouchy}, {Bonomo}, {Santerne}, {Moutou},
  {Deleuil}, {D{\'{\i}}az}, {Eggenberger}, {Ehrenreich}, {Gry}, {Guillot},
  {Havel}, {H{\'e}brard}, \& {Udry}}]{bouchy2011}
{Bouchy}, F., {Bonomo}, A.~S., {Santerne}, A., {et~al.} 2011, \aap, 533, A83

\bibitem[{{Bouchy} {et~al.}(2013){Bouchy}, {D{\'{\i}}az}, {H{\'e}brard},
  {Arnold}, {Boisse}, {Delfosse}, {Perruchot}, \& {Santerne}}]{bouchy2013}
{Bouchy}, F., {D{\'{\i}}az}, R.~F., {H{\'e}brard}, G., {et~al.} 2013, \aap,
  549, A49

\bibitem[{{Bouchy} {et~al.}(2009){Bouchy}, {H{\'e}brard}, {Udry}, {Delfosse},
  {Boisse}, {Desort}, {Bonfils}, {Eggenberger}, {Ehrenreich}, {Forveille},
  {Lagrange}, {Le Coroller}, {Lovis}, {Moutou}, {Pepe}, {Perrier}, {Pont},
  {Queloz}, {Santos}, {S{\'e}gransan}, \& {Vidal-Madjar}}]{bouchy2009}
{Bouchy}, F., {H{\'e}brard}, G., {Udry}, S., {et~al.} 2009, \aap, 505, 853

\bibitem[{{Bressan} {et~al.}(2012){Bressan}, {Marigo}, {Girardi}, {Salasnich},
  {Dal Cero}, {Rubele}, \& {Nanni}}]{bressan2012}
{Bressan}, A., {Marigo}, P., {Girardi}, L., {et~al.} 2012, \mnras, 427, 127

\bibitem[{{Bruntt} {et~al.}(2010{\natexlab{a}}){Bruntt}, {Bedding}, {Quirion},
  {Lo Curto}, {Carrier}, {Smalley}, {Dall}, {Arentoft}, {Bazot}, \&
  {Butler}}]{bruntt201023}
{Bruntt}, H., {Bedding}, T.~R., {Quirion}, P.-O., {et~al.} 2010{\natexlab{a}},
  \mnras, 405, 1907

\bibitem[{{Bruntt} {et~al.}(2010{\natexlab{b}}){Bruntt}, {Deleuil}, {Fridlund},
  {Alonso}, {Bouchy}, {Hatzes}, {Mayor}, {Moutou}, \& {Queloz}}]{bruntt2010c7}
{Bruntt}, H., {Deleuil}, M., {Fridlund}, M., {et~al.} 2010{\natexlab{b}}, \aap,
  519, A51

\bibitem[{{Chambers}(1999)}]{chambers1999}
{Chambers}, J.~E. 1999, \mnras, 304, 793

\bibitem[{{Cutri} {et~al.}(2003){Cutri}, {Skrutskie}, {van Dyk}, {Beichman},
  {Carpenter}, {Chester}, {Cambresy}, {Evans}, {Fowler}, {Gizis}, {Howard},
  {Huchra}, {Jarrett}, {Kopan}, {Kirkpatrick}, {Light}, {Marsh}, {McCallon},
  {Schneider}, {Stiening}, {Sykes}, {Weinberg}, {Wheaton}, {Wheelock}, \&
  {Zacarias}}]{cutri2003}
{Cutri}, R.~M., {Skrutskie}, M.~F., {van Dyk}, S., {et~al.} 2003, VizieR Online
  Data Catalog, 2246, 0

\bibitem[{{Dawson} {et~al.}(2014){Dawson}, {Johnson}, {Fabrycky},
  {Foreman-Mackey}, {Murray-Clay}, {Buchhave}, {Cargile}, {Clubb}, {Fulton},
  {Hebb}, {Howard}, {Huber}, {Shporer}, \& {Valenti}}]{dawson2014}
{Dawson}, R.~I., {Johnson}, J.~A., {Fabrycky}, D.~C., {et~al.} 2014, \apj, 791,
  89

\bibitem[{{Delisle} {et~al.}(2014){Delisle}, {Laskar}, \&
  {Correia}}]{delisle2014}
{Delisle}, J.-B., {Laskar}, J., \& {Correia}, A.~C.~M. 2014, \aap, 566, A137

\bibitem[{{D{\'{\i}}az} {et~al.}(2014){D{\'{\i}}az}, {Almenara}, {Santerne},
  {Moutou}, {Lethuillier}, \& {Deleuil}}]{diaz2014}
{D{\'{\i}}az}, R.~F., {Almenara}, J.~M., {Santerne}, A., {et~al.} 2014, \mnras,
  441, 983

\bibitem[{{Dotter} {et~al.}(2008){Dotter}, {Chaboyer}, {Jevremovi{\'c}},
  {Kostov}, {Baron}, \& {Ferguson}}]{dotter2008}
{Dotter}, A., {Chaboyer}, B., {Jevremovi{\'c}}, D., {et~al.} 2008, \apjs, 178,
  89

\bibitem[{{Everett} {et~al.}(2013){Everett}, {Howell}, {Silva}, \&
  {Szkody}}]{everett2013}
{Everett}, M.~E., {Howell}, S.~B., {Silva}, D.~R., \& {Szkody}, P. 2013, \apj,
  771, 107

\bibitem[{{Fabrycky} {et~al.}(2014){Fabrycky}, {Lissauer}, {Ragozzine}, {Rowe},
  {Steffen}, {Agol}, {Barclay}, {Batalha}, {Borucki}, {Ciardi}, {Ford},
  {Gautier}, {Geary}, {Holman}, {Jenkins}, {Li}, {Morehead}, {Morris},
  {Shporer}, {Smith}, {Still}, \& {Van Cleve}}]{fabrycky2014}
{Fabrycky}, D.~C., {Lissauer}, J.~J., {Ragozzine}, D., {et~al.} 2014, \apj,
  790, 146

\bibitem[{{Fitzpatrick}(1999)}]{fitzpatrick1999}
{Fitzpatrick}, E.~L. 1999, \pasp, 111, 63

\bibitem[{{Gaudi} \& {Winn}(2007)}]{gaudi2007}
{Gaudi}, B.~S. \& {Winn}, J.~N. 2007, \apj, 655, 550

\bibitem[{{H{\'e}brard} {et~al.}(2008){H{\'e}brard}, {Bouchy}, {Pont},
  {Loeillet}, {Rabus}, {Bonfils}, {Moutou}, {Boisse}, {Delfosse}, {Desort},
  {Eggenberger}, {Ehrenreich}, {Forveille}, {Lagrange}, {Lovis}, {Mayor},
  {Pepe}, {Perrier}, {Queloz}, {Santos}, {S{\'e}gransan}, {Udry}, \&
  {Vidal-Madjar}}]{hebrard2008}
{H{\'e}brard}, G., {Bouchy}, F., {Pont}, F., {et~al.} 2008, \aap, 488, 763

\bibitem[{{Holman} {et~al.}(2010){Holman}, {Fabrycky}, {Ragozzine}, {Ford},
  {Steffen}, {Welsh}, {Lissauer}, {Latham}, {Marcy}, {Walkowicz}, {Batalha},
  {Jenkins}, {Rowe}, {Cochran}, {Fressin}, {Torres}, {Buchhave}, {Sasselov},
  {Borucki}, {Koch}, {Basri}, {Brown}, {Caldwell}, {Charbonneau}, {Dunham},
  {Gautier}, {Geary}, {Gilliland}, {Haas}, {Howell}, {Ciardi}, {Endl},
  {Fischer}, {F{\"u}r{\'e}sz}, {Hartman}, {Isaacson}, {Johnson}, {MacQueen},
  {Moorhead}, {Morehead}, \& {Orosz}}]{holman2010}
{Holman}, M.~J., {Fabrycky}, D.~C., {Ragozzine}, D., {et~al.} 2010, Science,
  330, 51

\bibitem[{{Holman} \& {Murray}(2005)}]{holman2005}
{Holman}, M.~J. \& {Murray}, N.~W. 2005, Science, 307, 1288

\bibitem[{{Howard} {et~al.}(2010){Howard}, {Marcy}, {Johnson}, {Fischer},
  {Wright}, {Isaacson}, {Valenti}, {Anderson}, {Lin}, \& {Ida}}]{howard2010}
{Howard}, A.~W., {Marcy}, G.~W., {Johnson}, J.~A., {et~al.} 2010, Science, 330,
  653

\bibitem[{{Jenkins} {et~al.}(2010){Jenkins}, {Caldwell}, {Chandrasekaran},
  {Twicken}, {Bryson}, {Quintana}, {Clarke}, {Li}, {Allen}, {Tenenbaum}, {Wu},
  {Klaus}, {Middour}, {Cote}, {McCauliff}, {Girouard}, {Gunter}, {Wohler},
  {Sommers}, {Hall}, {Uddin}, {Wu}, {Bhavsar}, {Van Cleve}, {Pletcher},
  {Dotson}, {Haas}, {Gilliland}, {Koch}, \& {Borucki}}]{jenkins2010}
{Jenkins}, J.~M., {Caldwell}, D.~A., {Chandrasekaran}, H., {et~al.} 2010,
  \apjl, 713, L87

\bibitem[{{Jontof-Hutter} {et~al.}(2014){Jontof-Hutter}, {Lissauer}, {Rowe}, \&
  {Fabrycky}}]{jontof-hutter2014}
{Jontof-Hutter}, D., {Lissauer}, J.~J., {Rowe}, J.~F., \& {Fabrycky}, D.~C.
  2014, \apj, 785, 15

\bibitem[{{Kipping}(2010)}]{kipping2010}
{Kipping}, D.~M. 2010, \mnras, 408, 1758

\bibitem[{{Kipping}(2014)}]{kipping2014}
{Kipping}, D.~M. 2014, \mnras, 440, 2164

\bibitem[{{Lagarde} {et~al.}(2012){Lagarde}, {Decressin}, {Charbonnel},
  {Eggenberger}, {Ekstr{\"o}m}, \& {Palacios}}]{Lagarde2012}
{Lagarde}, N., {Decressin}, T., {Charbonnel}, C., {et~al.} 2012, \aap, 543,
  A108

\bibitem[{{Lega} {et~al.}(2013){Lega}, {Morbidelli}, \&
  {Nesvorn{\'y}}}]{lega2013}
{Lega}, E., {Morbidelli}, A., \& {Nesvorn{\'y}}, D. 2013, \mnras, 431, 3494

\bibitem[{{Lissauer} {et~al.}(2014){Lissauer}, {Marcy}, {Bryson}, {Rowe},
  {Jontof-Hutter}, {Agol}, {Borucki}, {Carter}, {Ford}, {Gilliland}, {Kolbl},
  {Star}, {Steffen}, \& {Torres}}]{lissauer2014}
{Lissauer}, J.~J., {Marcy}, G.~W., {Bryson}, S.~T., {et~al.} 2014, \apj, 784,
  44

\bibitem[{{Lissauer} {et~al.}(2012){Lissauer}, {Marcy}, {Rowe}, {Bryson},
  {Adams}, {Buchhave}, {Ciardi}, {Cochran}, {Fabrycky}, {Ford}, {Fressin},
  {Geary}, {Gilliland}, {Holman}, {Howell}, {Jenkins}, {Kinemuchi}, {Koch},
  {Morehead}, {Ragozzine}, {Seader}, {Tanenbaum}, {Torres}, \&
  {Twicken}}]{lissauer2012}
{Lissauer}, J.~J., {Marcy}, G.~W., {Rowe}, J.~F., {et~al.} 2012, \apj, 750, 112

\bibitem[{{Lissauer} {et~al.}(2011){Lissauer}, {Ragozzine}, {Fabrycky},
  {Steffen}, {Ford}, {Jenkins}, {Shporer}, {Holman}, {Rowe}, {Quintana},
  {Batalha}, {Borucki}, {Bryson}, {Caldwell}, {Carter}, {Ciardi}, {Dunham},
  {Fortney}, {Gautier}, {Howell}, {Koch}, {Latham}, {Marcy}, {Morehead}, \&
  {Sasselov}}]{lissauer2011}
{Lissauer}, J.~J., {Ragozzine}, D., {Fabrycky}, D.~C., {et~al.} 2011, \apjs,
  197, 8

\bibitem[{{Lithwick} \& {Wu}(2012)}]{lithwick2012}
{Lithwick}, Y. \& {Wu}, Y. 2012, \apjl, 756, L11

\bibitem[{{Mazeh} {et~al.}(2013){Mazeh}, {Nachmani}, {Holczer}, {Fabrycky},
  {Ford}, {Sanchis-Ojeda}, {Sokol}, {Rowe}, {Zucker}, {Agol}, {Carter},
  {Lissauer}, {Quintana}, {Ragozzine}, {Steffen}, \& {Welsh}}]{mazeh2013}
{Mazeh}, T., {Nachmani}, G., {Holczer}, T., {et~al.} 2013, \apjs, 208, 16

\bibitem[{{Mowlavi} {et~al.}(2012){Mowlavi}, {Eggenberger}, {Meynet},
  {Ekstr{\"o}m}, {Georgy}, {Maeder}, {Charbonnel}, \& {Eyer}}]{mowlavi2012}
{Mowlavi}, N., {Eggenberger}, P., {Meynet}, G., {et~al.} 2012, \aap, 541, A41

\bibitem[{{Nesvorn{\'y}} {et~al.}(2013){Nesvorn{\'y}}, {Kipping}, {Terrell},
  {Hartman}, {Bakos}, \& {Buchhave}}]{nesvorny2013}
{Nesvorn{\'y}}, D., {Kipping}, D., {Terrell}, D., {et~al.} 2013, \apj, 777, 3

\bibitem[{{Ofir} {et~al.}(2014){Ofir}, {Dreizler}, {Von Essen}, \&
  {Aharonson}}]{ofirtoulouse}
{Ofir}, A., {Dreizler}, S., {Von Essen}, C., \& {Aharonson}, O. 2014, in
  CoRoT3-KASC7 Symposium: The Space Photometry Revolution, ed. J. Ballot, \& R.
  A. Garcia, EPJ Web of Conferences, in press

\bibitem[{{Oshagh} {et~al.}(2013){Oshagh}, {Santos}, {Boisse}, {Bou{\'e}},
  {Montalto}, {Dumusque}, \& {Haghighipour}}]{oshagh2013}
{Oshagh}, M., {Santos}, N.~C., {Boisse}, I., {et~al.} 2013, \aap, 556, A19

\bibitem[{{Perruchot} {et~al.}(2008){Perruchot}, {Kohler}, {Bouchy}, {Richaud},
  {Richaud}, {Moreaux}, {Merzougui}, {Sottile}, {Hill}, {Knispel}, {Regal},
  {Meunier}, {Ilovaisky}, {Le Coroller}, {Gillet}, {Schmitt}, {Pepe}, {Fleury},
  {Sosnowska}, {Vors}, {M{\'e}gevand}, {Blanc}, {Carol}, {Point}, {Laloge}, \&
  {Brunel}}]{perruchot2008}
{Perruchot}, S., {Kohler}, D., {Bouchy}, F., {et~al.} 2008, in Society of
  Photo-Optical Instrumentation Engineers (SPIE) Conference Series, Vol. 7014,
  Society of Photo-Optical Instrumentation Engineers (SPIE) Conference Series

\bibitem[{{Pollacco} {et~al.}(2008){Pollacco}, {Skillen}, {Collier Cameron},
  {Loeillet}, {Stempels}, {Bouchy}, {Gibson}, {Hebb}, {H{\'e}brard}, {Joshi},
  {McDonald}, {Smalley}, {Smith}, {Street}, {Udry}, {West}, {Wilson},
  {Wheatley}, {Aigrain}, {Alsubai}, {Benn}, {Bruce}, {Christian}, {Clarkson},
  {Enoch}, {Evans}, {Fitzsimmons}, {Haswell}, {Hellier}, {Hickey}, {Hodgkin},
  {Horne}, {Hrudkov{\'a}}, {Irwin}, {Kane}, {Keenan}, {Lister}, {Maxted},
  {Mayor}, {Moutou}, {Norton}, {Osborne}, {Parley}, {Pont}, {Queloz}, {Ryans},
  \& {Simpson}}]{pollacco2008}
{Pollacco}, D., {Skillen}, I., {Collier Cameron}, A., {et~al.} 2008, \mnras,
  385, 1576

\bibitem[{{Press} \& {Rybicki}(1989)}]{press1989}
{Press}, W.~H. \& {Rybicki}, G.~B. 1989, \apj, 338, 277

\bibitem[{{Queloz} {et~al.}(2001){Queloz}, {Henry}, {Sivan}, {Baliunas},
  {Beuzit}, {Donahue}, {Mayor}, {Naef}, {Perrier}, \& {Udry}}]{queloz2001}
{Queloz}, D., {Henry}, G.~W., {Sivan}, J.~P., {et~al.} 2001, \aap, 379, 279

\bibitem[{{Rowe} {et~al.}(2014){Rowe}, {Bryson}, {Marcy}, {Lissauer},
  {Jontof-Hutter}, {Mullally}, {Gilliland}, {Issacson}, {Ford}, {Howell},
  {Borucki}, {Haas}, {Huber}, {Steffen}, {Thompson}, {Quintana}, {Barclay},
  {Still}, {Fortney}, {Gautier}, {Hunter}, {Caldwell}, {Ciardi}, {Devore},
  {Cochran}, {Jenkins}, {Agol}, {Carter}, \& {Geary}}]{rowe2014}
{Rowe}, J.~F., {Bryson}, S.~T., {Marcy}, G.~W., {et~al.} 2014, \apj, 784, 45

\bibitem[{{Santerne} {et~al.}(2012){Santerne}, {D{\'{\i}}az}, {Moutou},
  {Bouchy}, {H{\'e}brard}, {Almenara}, {Bonomo}, {Deleuil}, \&
  {Santos}}]{santerne2012}
{Santerne}, A., {D{\'{\i}}az}, R.~F., {Moutou}, C., {et~al.} 2012, \aap, 545,
  A76

\bibitem[{{Santerne} {et~al.}(2014){Santerne}, {H{\'e}brard}, {Deleuil},
  {Havel}, {Correia}, {Almenara}, {Alonso}, {Arnold}, {Barros}, {Behrend},
  {Bernasconi}, {Boisse}, {Bonomo}, {Bouchy}, {Bruno}, {Damiani},
  {D{\'{\i}}az}, {Gravallon}, {Guillot}, {Labrevoir}, {Montagnier}, {Moutou},
  {Rinner}, {Santos}, {Abe}, {Audejean}, {Bendjoya}, {Gillier}, {Gregorio},
  {Martinez}, {Michelet}, {Montaigut}, {Poncy}, {Rivet}, {Rousseau}, {Roy},
  {Suarez}, {Vanhuysse}, \& {Verilhac}}]{santerne2014}
{Santerne}, A., {H{\'e}brard}, G., {Deleuil}, M., {et~al.} 2014, \aap, 571, A37

\bibitem[{{Skrutskie} {et~al.}(2006){Skrutskie}, {Cutri}, {Stiening},
  {Weinberg}, {Schneider}, {Carpenter}, {Beichman}, {Capps}, {Chester},
  {Elias}, {Huchra}, {Liebert}, {Lonsdale}, {Monet}, {Price}, {Seitzer},
  {Jarrett}, {Kirkpatrick}, {Gizis}, {Howard}, {Evans}, {Fowler}, {Fullmer},
  {Hurt}, {Light}, {Kopan}, {Marsh}, {McCallon}, {Tam}, {Van Dyk}, \&
  {Wheelock}}]{sktutskie2006}
{Skrutskie}, M.~F., {Cutri}, R.~M., {Stiening}, R., {et~al.} 2006, \aj, 131,
  1163

\bibitem[{{Steffen} {et~al.}(2010){Steffen}, {Batalha}, {Borucki}, {Buchhave},
  {Caldwell}, {Cochran}, {Endl}, {Fabrycky}, {Fressin}, {Ford}, {Fortney},
  {Haas}, {Holman}, {Howell}, {Isaacson}, {Jenkins}, {Koch}, {Latham},
  {Lissauer}, {Moorhead}, {Morehead}, {Marcy}, {MacQueen}, {Quinn},
  {Ragozzine}, {Rowe}, {Sasselov}, {Seager}, {Torres}, \&
  {Welsh}}]{steffen2010}
{Steffen}, J.~H., {Batalha}, N.~M., {Borucki}, W.~J., {et~al.} 2010, \apj, 725,
  1226

\bibitem[{{Weiss} \& {Marcy}(2014)}]{weiss2014}
{Weiss}, L.~M. \& {Marcy}, G.~W. 2014, \apjl, 783, L6

\bibitem[{{Winn}(2010)}]{winn2010}
{Winn}, J.~N. 2010, arXiv:1001.2010

\bibitem[{{Wright} {et~al.}(2010){Wright}, {Eisenhardt}, {Mainzer}, {Ressler},
  {Cutri}, {Jarrett}, {Kirkpatrick}, {Padgett}, {McMillan}, {Skrutskie},
  {Stanford}, {Cohen}, {Walker}, {Mather}, {Leisawitz}, {Gautier}, {McLean},
  {Benford}, {Lonsdale}, {Blain}, {Mendez}, {Irace}, {Duval}, {Liu}, {Royer},
  {Heinrichsen}, {Howard}, {Shannon}, {Kendall}, {Walsh}, {Larsen}, {Cardon},
  {Schick}, {Schwalm}, {Abid}, {Fabinsky}, {Naes}, \& {Tsai}}]{wright2010}
{Wright}, E.~L., {Eisenhardt}, P.~R.~M., {Mainzer}, A.~K., {et~al.} 2010, \aj,
  140, 1868

\bibitem[{{Wright} {et~al.}(2011){Wright}, {Fakhouri}, {Marcy}, {Han}, {Feng},
  {Johnson}, {Howard}, {Fischer}, {Valenti}, {Anderson}, \&
  {Piskunov}}]{wright2011}
{Wright}, J.~T., {Fakhouri}, O., {Marcy}, G.~W., {et~al.} 2011, \pasp, 123, 412

\end{thebibliography}

\Online

\clearpage

\appendix

\section{Additional figures and tables}

\begin{table}[htbp]
\centering
\caption{\label{tabttvs} Transit-timing variations for planet b.}
\begin{tabular}{|c|c|c|c|}
\hline\hline
Epoch & Mid-transit time  & TTV & Uncertainty \\
    & [BJD - 2450000] & [min] & [min] \\
\hline
0   & 4978.82211 &  -2.9 & 2.7  \\
1   & 4997.62519 &   7.5 & 3.0  \\
3   & 5035.19749 & -20.6 & 2.5  \\
4   & 5054.00955 &   2.6 & 3.5  \\
5   & 5072.80069 &  -4.2 & 2.9  \\
7   & 5110.40214 &   9.6 & 2.8  \\
8   & 5129.19445 &   4.5 & 3.0  \\
9   & 5147.98545 &  -2.6 & 2.6  \\
10  & 5166.78207 &  -1.6 & 2.5  \\
11  & 5185.57076 & -12.0 & 2.4  \\
12  & 5204.37575 &   1.1 & 2.2  \\
14  & 5241.95112 & -22.6 & 2.2  \\
15  & 5260.76796 &   7.6 & 2.2  \\
16  & 5279.56177 &   4.6 & 2.4  \\
17  & 5298.35371 &  -1.2 & 2.4  \\
18  & 5317.15385 &   4.9 & 2.4  \\
19  & 5335.94633 &   0.0 & 2.3  \\
20  & 5354.74575 &   5.1 & 2.3  \\
21  & 5373.52714 & -15.8 & 2.3  \\
22  & 5392.32105 & -18.7 & 2.3  \\
23  & 5411.13384 &   5.6 & 2.2  \\
24  & 5429.92650 &   0.9 & 2.2  \\
25  & 5448.71835 &  -4.9 & 2.2  \\
26  & 5467.52116 &   5.0 & 2.5  \\
27  & 5486.31647 &   4.2 & 2.4  \\
28  & 5505.10995 &   0.7 & 2.4  \\
30  & 5542.68973 & -16.7 & 2.4  \\
32  & 5580.29000 &  -4.5 & 2.2  \\
33  & 5599.08156 & -10.8 & 2.3  \\
34  & 5617.89305 &  11.7 & 2.2  \\
36  & 5655.47836 &   2.3 & 2.4  \\
37  & 5674.27106 &  -2.3 & 2.4  \\
38  & 5693.06165 &  -9.9 & 2.5  \\
39  & 5711.86685 &   3.4 & 2.3  \\
41  & 5749.44394 & -17.8 & 2.2  \\
42  & 5768.25766 &   7.9 & 2.2  \\
43  & 5787.05285 &   6.8 & 2.2  \\
44  & 5805.84231 &  -2.4 & 2.2  \\
45  & 5824.64242 &   3.6 & 2.2  \\
46  & 5843.43402 &  -2.6 & 2.4  \\
47  & 5862.23795 &   9.0 & 2.5  \\
48  & 5881.02176 &  -8.5 & 2.4  \\
49  & 5899.80911 & -20.8 & 2.4  \\
50  & 5918.62126 &   2.6 & 2.4  \\
51  & 5937.41712 &   2.5 & 2.2  \\
52  & 5956.20759 &  -5.3 & 2.9  \\
53  & 5975.01304 &   8.4 & 2.1  \\
55  & 6012.59970 &   1.0 & 2.2  \\
56  & 6031.38973 &  -7.4 & 2.4  \\
57  & 6050.18187 & -12.8 & 2.3  \\
58  & 6068.99024 &   5.1 & 2.4  \\
59  & 6087.77696 &  -8.1 & 2.4  \\
62  & 6144.17234 &   2.9 & 2.3  \\
63  & 6162.96539 &  -1.2 & 2.3  \\
64  & 6181.75998 &  -3.1 & 2.2  \\
65  & 6200.55215 &  -8.5 & 2.3  \\
66  & 6219.35264 &  -1.9 & 2.5  \\
68  & 6256.93344 & -17.8 & 2.4  \\
69  & 6275.74812 &   9.3 & 2.4  \\
70  & 6294.54117 &   5.2 & 2.5  \\
72  & 6332.13746 &  11.6 & 2.2  \\
73  & 6350.92542 &   0.2 & 2.2  \\
74  & 6369.72444 &   4.6 & 2.2  \\
75  & 6388.51087 &  -9.0 & 2.2  \\
76  & 6407.30314 & -14.2 & 2.4  \\
\hline
\end{tabular}
\label{ttvb}
\end{table}

\begin{table}[htbp]
\centering
\caption{\label{tabttvs2} Transit-timing variations for planet c.}
\begin{tabular}{|c|c|c|c|}
\hline\hline
Epoch & Mid-transit time  & TTV & Uncertainty \\
    & [BJD - 2450000] & [min] & [min] \\
\hline
0   &  4968.63008   &    -3.5 &  1.3 \\
1   &  5019.42134   &    -2.2 &  1.3 \\
2   &  5070.21489   &     2.3 &  1.3 \\
3   &  5121.00355   &    -0.1 &  1.3 \\
4   &  5171.79086   &    -4.5 &  1.4 \\
6   &  5273.37520   &     0.6 &  1.2 \\
7   &  5324.16706   &     2.8 &  1.3 \\
8   &  5374.95617   &     0.9 &  1.2 \\
9   &  5425.74606   &     0.3 &  1.2 \\
10  &  5476.53664   &     0.6 &  1.3 \\
11  &  5527.32559   &   -1.5 &  1.3 \\
12  &  5578.11883   &     2.6 &  1.2 \\
13  &  5628.90845   &     1.6 &  1.2 \\
14  &  5679.69704   &   -1.0 &  1.3 \\
16  &  5781.27846   &   0.0 &  1.2 \\
17  &  5832.06706   &    -2.6 &  1.2 \\
18  &  5882.85959   &     0.5 &  1.3 \\
19  &  5933.65090   &     1.9 &  1.2 \\
20  &  5984.43899   &    -1.4 &  1.2 \\
21  &  6035.23056   &     0.3 &  1.3 \\
22  &  6086.02001   &    -1.0 &  1.3 \\
23  &  6136.81231   &     1.7 &  1.2 \\
24  &  6187.60050   &    -1.4 &  1.2 \\
26  &  6289.18225   &     0.0 &  1.3 \\
27  &  6339.97047   &    -3.1 &  1.2 \\
28  &  6390.76238   &    -0.8 &  1.6 \\
\hline
\end{tabular}
\label{ttvc}
\end{table}

\clearpage

\begin{table*}[htbp]
\caption{\label{priors} Prior distributions used in the combined fit with \texttt{PASTIS}. $\mathcal{U}(a, b)$ stands for a uniform distribution between $a$ and $b$; $\mathcal{N}(\mu, \sigma)$ indicates a normal distribution with mean $\mu$ and standard deviation $\sigma$; $\mathcal{N_A}(\mu, \sigma_-, \sigma_+)$, stands for an asymmetric normal with mean $\mu$, right width $\sigma_+$ and left width $\sigma_-$; $\mathcal{S}(a, b)$ represents a sine distribution between $a$ and $b$; finally, $\mathcal{J}(a, b)$ means a Jeffreys distribution between $a$ and $b$.}
\centering
\begin{tabular}{lll}
\hline\hline
\multicolumn{3}{l}{\emph{Stellar parameters}} \smallskip\\
Effective temperature \teff\ [K] & $\mathcal{N}(6169, 100)$ \\
Metallicity \feh\    & $\mathcal{N}(-0.04, 0.10)$ \\
Stellar density $\rho_{\star}$ [$\rho_\odot$]  (Dartmouth) & $\mathcal{N_A}(0.25, 0.08, 0.29)$  \\
Stellar density $\rho_{\star}$ [$\rho_\odot$]  (PARSEC) & $\mathcal{N_A}(0.25, 0.08, 0.30)$\\
Stellar density $\rho_{\star}$ [$\rho_\odot$]  (StarEvol) & $\mathcal{N_A}(0.23, 0.07, 0.25)$\\
Stellar density $\rho_{\star}$ [$\rho_\odot$]  (Geneva) & $ \mathcal{N_A}(0.24, 0.07, 0.30)$\\
Quadratic limb-darkening coefficient $u_a$     & $\mathcal{U}(-0.5, 1.2)$\\
Quadratic limb-darkening coefficient $u_b$     & $\mathcal{U}(-0.5, 1.2)$ \\
Stellar RV linear drift [\ms yr$^{-1}$]                 & $\mathcal{U}(-0.001, 0.001)$ \smallskip\\

\multicolumn{1}{l}{\emph{Planet parameters}} & \emph{Kepler-117 b} & \emph{Kepler-117 c} \smallskip\\
Orbital period $P$ [days]                   & $\mathcal{N}(18.795921, 1.3 \times 10^{-5})$ & $\mathcal{N}(50.790391, 2.3\times 10^{-5})$ \\
Primary transit epoch $T_{0}$ [BJD-2450000] & $\mathcal{N}(4978.82194, 4.7\times 10^{-4})$ & $\mathcal{N}(4968.63195, 3.1\times 10^{-4})$\\
Orbital eccentricity $e$                    & $\mathcal{U}(0, 1)$ & $\mathcal{U}(0, 1)$ \\
Argument of periastron $\omega$ [deg]       & $\mathcal{U}(0, 360)$ & $\mathcal{U}(0, 360)$\\
Orbital inclination $i$ [deg]               & $\mathcal{S}(50, 90)$ &            $\mathcal{S}(89, 91)$ \\
Longitude of the ascending node $\Omega$ [deg]       & $\mathcal{U}(135, 225)$ & - \\
Radius ratio $R_p/R_\star$   & $\mathcal{J}(0.01, 0.50)$ & $\mathcal{J}(0.01, 0.50)$ \\
Radial velocity semi-amplitude $K$ [\ms]    & $\mathcal{U}(0.0, 0.5)$ & $\mathcal{U}(0.0, 0.5)$ \smallskip\\

\multicolumn{3}{l}{\emph{System parameters}} \smallskip\\
Distance [pc]  & $\mathcal{U}(0, 1\times10^4)$  \\
Interstellar extinction $E(B - V)$ & $\mathcal{U}(0, 3)$\\
Systemic velocity (BJD 2456355), $V_{r}$ [\kms] & $\mathcal{U}(-13.1, -12.6)$ \smallskip\\

\multicolumn{3}{l}{\emph{Instrumental parameters}} \smallskip\\
Kepler jitter (LC) [ppm] & $\mathcal{U}(0, 0.0007)$\\
Kepler offset (LC) [ppm] & $\mathcal{U}(0.99, 1.01)$\\
Kepler jitter (SC) [ppm] & $\mathcal{U}(0, 0.0004)$\\
Kepler offset (SC) [ppm] & $\mathcal{U}(0.99, 1.01)$\\
TTV jitter, planet b [min]          & $\mathcal{U}(0, 9)$\\
TTV jitter, planet c [min]          & $\mathcal{U}(0, 4)$  \\
SOPHIE jitter [\ms]        & $\mathcal{U}(0, 2)$\\
SED jitter [mags]          & $\mathcal{U}(0, 1)$ \\
\\
\hline
\end{tabular}
\end{table*}

\begin{table*}[htbp]
\begin{center}{
\caption{\label {tabmag} Magnitudes of the Kepler-117 system.}
\begin{tabular}{|c|c|c|}
\hline\hline
Filter-Band    &    mag  &    error \\
\hline
Johnson-B $^a$ & 15.056   &  0.029	\\ 
Johnson-V $^a$ & 14.476   &  0.027 \\
SDSS-G $^a$	  & 14.688   &  0.032\\
SDSS-R $^a$	  & 14.36    &  0.026\\
SDSS-I $^a$	  & 14.227   &  0.066 \\
2MASS-J $^b$   & 13.324   &  0.026\\
2MASS-H $^b$   & 12.988  &   0.031\\
2MASS-Ks $^b$  & 13.011   &  0.031\\
WISE-W1 $^c$   & 12.946   &  0.024 \\	
WISE-W2 $^c$   & 12.992   &  0.025 \\
\hline
\end{tabular}}
\end{center}
\begin{list}{}{}
\item $^a$ APASS (http://www.aavso.org/apass); $^b$ 2MASS \citep{sktutskie2006,cutri2003}; $^c$ WISE \citep{wright2010}.
\end{list}
\end{table*}

\begin{figure*}
\center
\includegraphics[scale = 0.45]{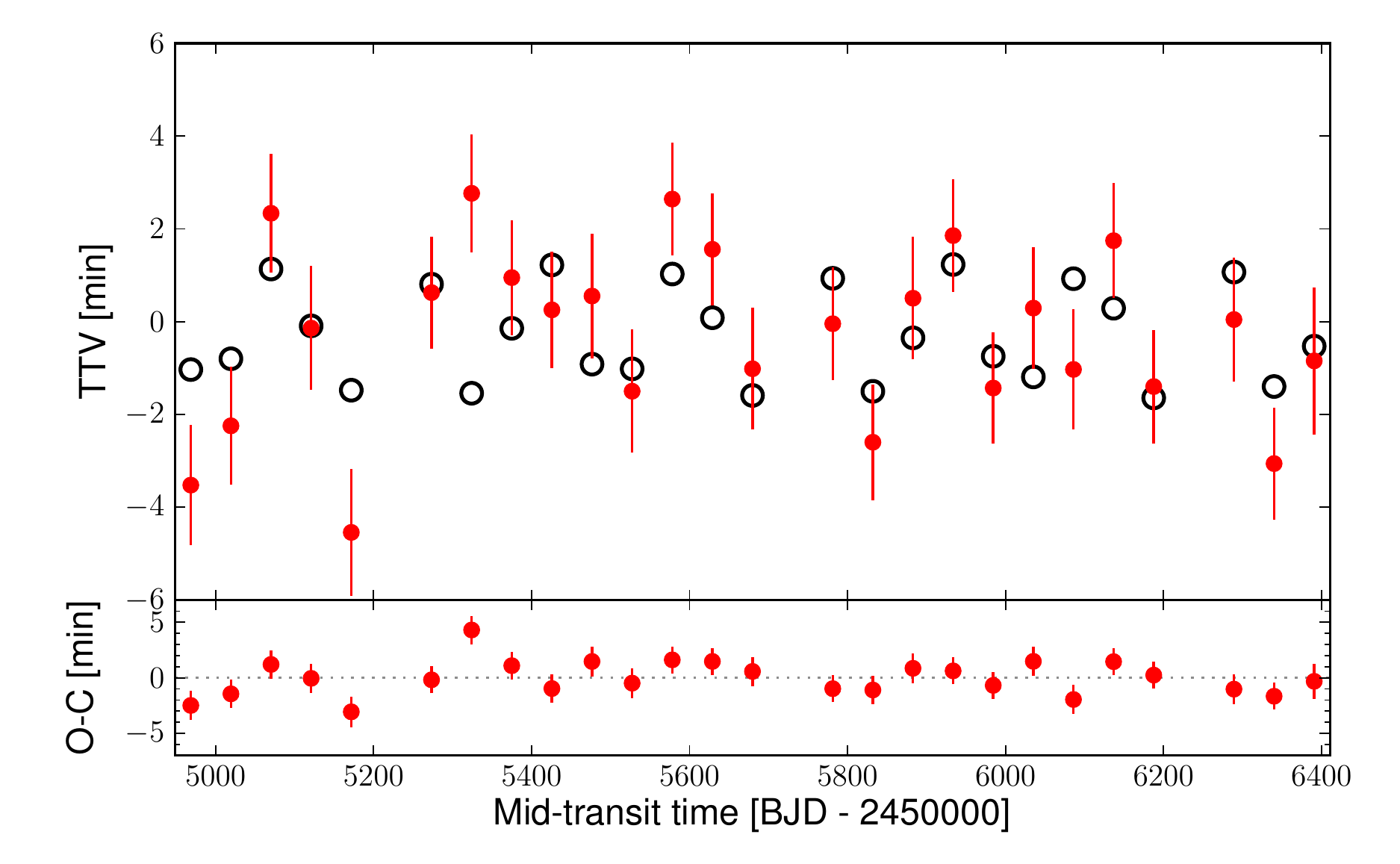}
\includegraphics[scale = 0.45]{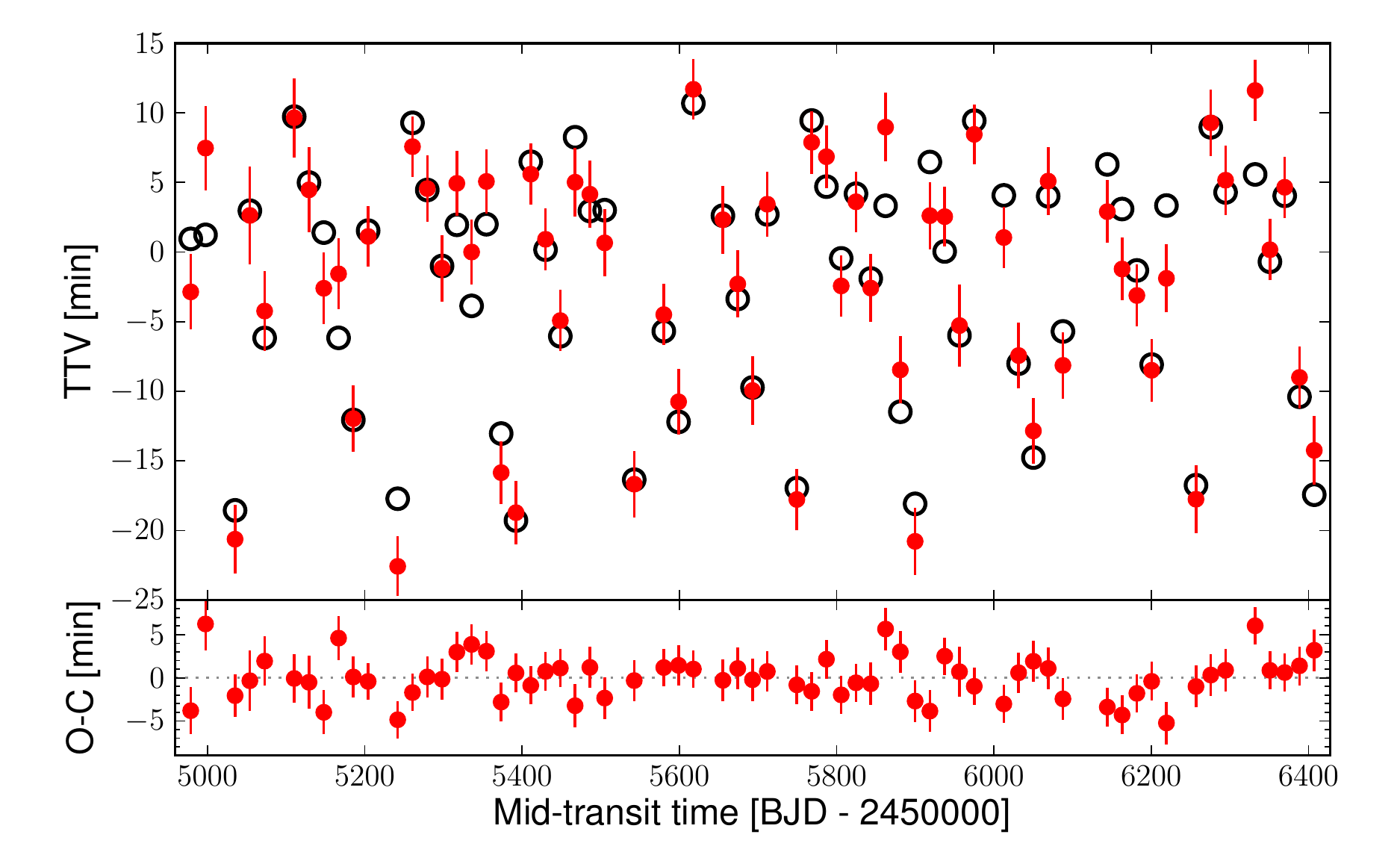}
\caption{TTVs of planet b (left) and c (right) as a function of time. In red the data, in black the fit. The lower panels show the residuals.}
\label{ttvss}
\end{figure*}

\begin{figure*}[htbp]
\centering
\includegraphics[scale = 0.8]{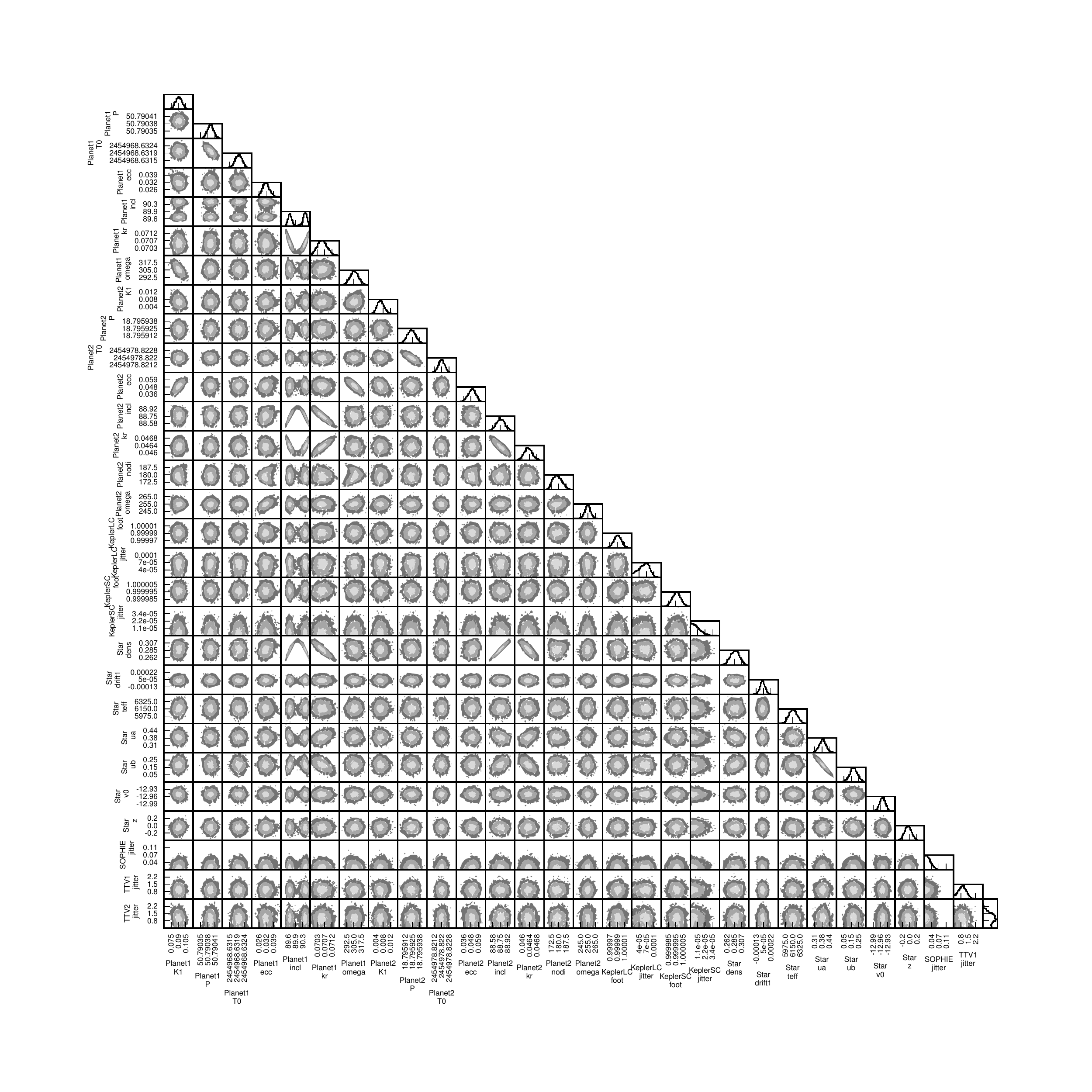}
\caption{Pyramid of the combined posterior distributions.}
\label{pyramid}
\end{figure*}

\end{document}